\documentclass[revtex4]{emulateapj}
\usepackage[]{graphicx}
\usepackage{subfigure}
\usepackage{aas_macros}
\usepackage{natbib}
\usepackage{amsmath}
\usepackage{hyperref}

\begin{document}

\title{Optical and near-infrared observations of SN 2013dx associated with GRB 130702A}
\author{
V.~L.~Toy\altaffilmark{1}, 
S.~B.~Cenko\altaffilmark{2}$^{,}$\altaffilmark{3}, 
J.~M.~Silverman\altaffilmark{4}$^{,}$\altaffilmark{5}, 
N.~R.~Butler\altaffilmark{6},
A.~Cucchiara\altaffilmark{7},
A.~M.~Watson\altaffilmark{8}, 
D.~Bersier\altaffilmark{9},
D.~A.~Perley\altaffilmark{10}$^{,}$\altaffilmark{11}, 
R.~Margutti\altaffilmark{12},
E.~Bellm\altaffilmark{10},
J.~S.~Bloom\altaffilmark{13},
Y.~Cao\altaffilmark{10},
J.~I.~Capone\altaffilmark{1},
K.~I.~Clubb\altaffilmark{13},
A.~Corsi\altaffilmark{14},
A.~De~Cia\altaffilmark{15},
J.~A.~de Diego\altaffilmark{8},
A.~V.~Filippenko\altaffilmark{13},
O.~D.~Fox\altaffilmark{13},
A.~Gal-Yam\altaffilmark{15},
N.~Gehrels\altaffilmark{2},
L.~Georgiev\altaffilmark{8},
J.~J.~Gonz\'{a}lez\altaffilmark{8},
M.~M.~Kasliwal\altaffilmark{16},
P.~L.~Kelly\altaffilmark{13},
S.~R.~Kulkarni\altaffilmark{10},
A.~S.~Kutyrev\altaffilmark{2},
W.~H.~Lee\altaffilmark{8},
J.~X.~Prochaska\altaffilmark{17},
E.~Ramirez-Ruiz\altaffilmark{17},
M.~G.~Richer\altaffilmark{8},
C.~Rom\'{a}n-Z\'u\~niga\altaffilmark{8},
L.~Singer\altaffilmark{18}$^{,}$\altaffilmark{7},
D.~Stern\altaffilmark{19},
E.~Troja\altaffilmark{2}$^{,}$\altaffilmark{1}, and
S.~Veilleux\altaffilmark{1}$^{,}$\altaffilmark{3}}

\altaffiltext{1} {Department of Astronomy, University of Maryland, College Park, MD 20742, USA}
\altaffiltext{2} {NASA, Goddard Space Flight Center, Greenbelt, MD 20771, USA}
\altaffiltext{3} {Joint Space-Science Institute, University of Maryland, College Park, MD 20742, USA}
\altaffiltext{4} {Department of Astronomy, University of Texas at Austin, Austin, TX 78712, USA}
\altaffiltext{5} {NSF Astronomy and Astrophysics Postdoctoral Fellow}
\altaffiltext{6} {School of Earth \& Space Exploration, Arizona State University, AZ 85287, USA}
\altaffiltext{7} {NASA Postdoctoral Program Fellow, Goddard Space Flight Center, Greenbelt, MD 20771, USA}
\altaffiltext{8} {Instituto de Astronom\'{i}a, Universidad Nacional Auto\'{n}oma de M\'{e}xico, Apartado Postal 106, 22800 Ensenada, Baja California, M\'{e}xico}
\altaffiltext{9} {Astrophysics Research Institute, Liverpool John Moores University, Liverpool L3 5RF, UK}
\altaffiltext{10} {Department of Astronomy, California Institute of Technology, MC 249-17, 1200 East California Blvd, Pasadena CA 91125, USA}
\altaffiltext{11} {Hubble Fellow}
\altaffiltext{12} {Harvard-Smithsonian Center for Astrophysics, 60 Garden St., Cambridge, MA 02138, USA}
\altaffiltext{13} {Department of Astronomy, University of California, Berkeley, CA 94720-3411, USA}
\altaffiltext{14} {Department of Physics, Texas Tech University, Box 41051, Lubbock, TX 79409-1051, USA}
\altaffiltext{15} {Department of Particle Physics and Astrophysics, Weizmann Institute of Science, Rehovot 7610001, Israel}
\altaffiltext{16} {Observatories of the Carnegie Institution for Science, 813 Santa Barbara Street, Pasadena CA 91101, USA}
\altaffiltext{17} {Department of Astronomy and Astrophysics, UCO/Lick Observatory, University of California, 1156 High Street, Santa Cruz, CA 95064, USA}
\altaffiltext{18} {LIGO Laboratory, California Institute of Technology, Pasadena, CA 91125, USA}
\altaffiltext{19} {Jet Propulsion Laboratory, California Institute of Technology, Pasadena, CA 91109, USA}

\begin{abstract}
We present optical and near-infrared light curves and optical spectra of SN 2013dx,
associated with the nearby (redshift 0.145) gamma-ray burst GRB\,130702A.  
The prompt isotropic gamma-ray energy released from GRB\,130702A
is measured to be $E_{\gamma,\mathrm{iso}} = 6.4_{-1.0}^{+1.3} \times 10^{50}$\,erg 
(1\,keV to 10\,MeV in the rest frame), placing it intermediate between low-luminosity GRBs 
like GRB\,980425/SN 1998bw and the broader cosmological population.  
We compare the observed $g^{\prime} r^{\prime}
i^{\prime} z^{\prime}$ light curves of SN 2013dx to a SN 1998bw 
template, finding that SN 2013dx evolves $\sim20$\% faster (steeper rise time), with a 
comparable peak luminosity.  Spectroscopically, SN 2013dx resembles other
broad-lined Type Ic supernovae, both associated with (SN 2006aj and SN 1998bw)
and lacking (SN 1997ef, SN 2007I, and SN 2010ah) gamma-ray emission, with 
photospheric velocities around peak of $\sim$\,21,000\,km\,s$^{-1}$.  We 
construct a quasi-bolometric ($g^{\prime} r^{\prime} i^{\prime} z^{\prime} y J$)
light curve for SN 2013dx, only the fifth GRB-associated SN with extensive 
near-infrared coverage and the third with a bolometric light curve extending beyond 
$\Delta t > 40$\,d.   Together with the measured photospheric velocity, we
derive basic explosion parameters using simple analytic models.  We infer
a $^{56}$Ni mass of $M_{\mathrm{Ni}} = 0.37\pm 0.01$\,M$_{\odot}$, an ejecta 
mass of $M_{\mathrm{ej}} = 3.1 \pm 0.1$\,M$_{\odot}$, and a kinetic energy of 
$E_{\mathrm{K}} = (8.2 \pm 0.43) \times 10^{51}$\,erg (statistical uncertainties
only), consistent with previous GRB-associated supernovae.  When considering the ensemble
population of GRB-associated supernovae, we find no correlation between the mass of 
synthesized $^{56}$Ni and high-energy properties, despite clear predictions from
numerical simulations that $M_{\mathrm{Ni}}$ should correlate with the degree
of asymmetry.  On the other hand, $M_{\mathrm{Ni}}$ clearly correlates with 
the kinetic energy of the supernova ejecta across a wide range of core-collapse
events.
\end{abstract}

%===================================================
% Keywords
%===================================================

%%%%%%%%%%%%%%%%%%%%%%%%%%%%%%%%%%%%%%%%%%%%%%

\section{Introduction}

The evidence for the association between long-duration gamma-ray bursts (GRBs) and the death of massive stars has been steadily growing over the last two decades (see \citealt{Woosley:2006} and \citealt{Hjorth:2012} for reviews).  The first direct evidence of this link was a spatially and temporally coincident supernova (SN), SN 1998bw, with GRB\,980425 at redshift $z=0.0085$ \citep{Galama:1998, Iwamoto:1998, Kulkarni:1998}.  Since SN 1998bw, there have been a number of spectroscopically confirmed supernovae (SNe) associated with GRBs (Table \ref{tab:GRBSNref}).

%%%%%%
%GRB\,030329/SN 2003dh \citep{Stanek:2003,mgs+03}, 
%GRB\,031203/SN 2003lw \citep{Malesani:2004,gmf+04,thw+04},
%GRB\,060218/SN 2006aj \citep{Campana:2006,Modjaz:2006,mha+06,fkz+06,sjf+06,pmm+06,kmb+07},
%GRB\,091127/SN 2009nz \citep{cbp+10,bch+11},
%GRB\,100316D/SN 2010bh \citep{Starling:2011,bps+12,ogs+12,Cano:2011,Chornock:2010},
%GRB\,120422A/SN 2012bx \citep{Melandri:2012,Schulze:2014}, 
%GRB\,130427A/SN 2013cq \citep{xdl+13,ltf+14,mpd+14,pcc+14}, and
%GRB\,140606B/iPTF14bfu \citep{cdp+15}.
%%%%%%

\begin{deluxetable*}{ll}
\tabletypesize{\footnotesize}
\tablecolumns{2}
\tablewidth{0pt}
\tablecaption{GRB-SN References}
\tablehead{
\colhead{GRB-SN} & \colhead{References}
}
\startdata
GRB 980425/SN 1998bw	& \citet{Galama:1998, Iwamoto:1998, Kulkarni:1998}\\
GRB 030329/SN 2003dh	& \citet{Stanek:2003,mgs+03}\\
GRB 031203/SN 2003lw     & \citet{Malesani:2004,gmf+04,thw+04}\\
GRB 060218/SN 2006aj	& \citet{Campana:2006,Modjaz:2006,mha+06, fkz+06}\\
					& \citet{sjf+06,pmm+06,kmb+07}\\
GRB 091127/SN 2009nz     & \citet{cbp+10,bch+11}\\
GRB 100316D/SN 2010bh  & \citet{Starling:2011,bps+12,ogs+12}\\
					& \citet{Cano:2011,Chornock:2010} \\
GRB 120422A/SN 2012bz  & \citet{Melandri:2012,Schulze:2014}\\
GRB 130427A/SN 2013cq  & \citet{xdl+13,ltf+14,mpd+14,pcc+14} \\
GRB 130702A/SN 2013dx  & This work; \citet{DElia:2015}\\
GRB 140606B/iPTF14bfu	& \citet{cdp+15}
\enddata
\tablecomments{We do not include the recent detection of GRB 111209A/SN 2011kl 
associated with a superluminous SN \citep{Greiner:2015} because it is believed to be 
powered by a magnetar and not solely powered by $^{56}$Ni.}
\label{tab:GRBSNref}
\end{deluxetable*}

While most, if not all low-$z$ long-duration GRBs appear to be accompanied by SNe 
(the exceptions being GRB 060614 and GRB 060505; \citealt{Fynbo:2006, Gal-Yam:2006, DellaValle:2006}), 
only a small fraction of core-collapse explosions are capable of generating relativistic ejecta
 \citep{bkf+03,Soderberg:2010,bdg+14}.  
Even when limited to the specific subtype of SNe 
associated with GRBs, the broad-lined Type Ic SNe, those with and without relativistic 
ejecta appear to be indistinguishable based on their light curves (e.g., \citealt{Drout:2011}). 
However, spectra of the host galaxies reveal that 
GRB-SNe prefer more metal-poor environments than Type Ic-BL SNe without associated 
GRBs \citep{Modjaz:2008, Graham:2013}.

Furthermore, within the GRB population, there is a considerable
diversity in the observed prompt gamma-ray energies spanning six orders of magnitude, $E_{\gamma,\mathrm{iso}} = 10^{48}$--$10^{54}$\,erg.  It has been suggested that low-luminosity GRBs ($E_{\gamma,\mathrm{iso}} \lesssim 10^{49}$\,erg) have ``failed" jets that cannot pierce their stellar envelope and instead dissipate energy into the star to create relativistic shock breakout \citep{Bromberg:2011,mms+14,n15}.  But 
despite their very different appearance at high energies, as of yet there is no
clear distinction between SNe associated with low-luminosity GRBs (e.g., SN 1998bw)
and the larger (observed) cosmological population (e.g., SN 2003dh, SN 2013cq). 

With still only a handful of well-observed examples, each new nearby GRB affords 
a unique opportunity to understand the \textit{central engine} powering these outflows.
In particular, we can probe the progenitor from two different angles by studying the SN simultaneously with the GRB.  Here we present observations of SN 2013dx associated
with GRB\,130702A.  At $z = 0.145$, SN 2013dx is sufficiently nearby to enable
a detailed photometric and spectroscopic study of the SN evolution.  Furthermore,
with $E_{\gamma,\mathrm{iso}} = 6.4 \times 10^{50}$\,erg (for 1\,keV to 10\,MeV in the rest frame), 
the prompt-emission properties place GRB\,130702A between most low-luminosity GRBs and the 
more energetic cosmological population.

Throughout this paper we use the convention $F_{\nu}(t) \propto \nu^{-\beta}t^{-\alpha}$ and photon index $\Gamma = \beta + 1$.  We assume a $\Lambda$CDM model with H$_0 = 69.6 \text{ km } \text{s}^{-1} \text{ Mpc}^{-1}$, $\Omega_m = 0.286$, and $\Omega_{\Lambda} = 0.714$ (\citealt{Bennett:2014}).  All photometry is in the AB system \citep{og83}, and quoted uncertainties are 1$\sigma$ (68\%) confidence intervals unless otherwise noted. Dates and times are UT in all cases.

\begin{figure*}[ht]
\centering
\includegraphics[width=7in]{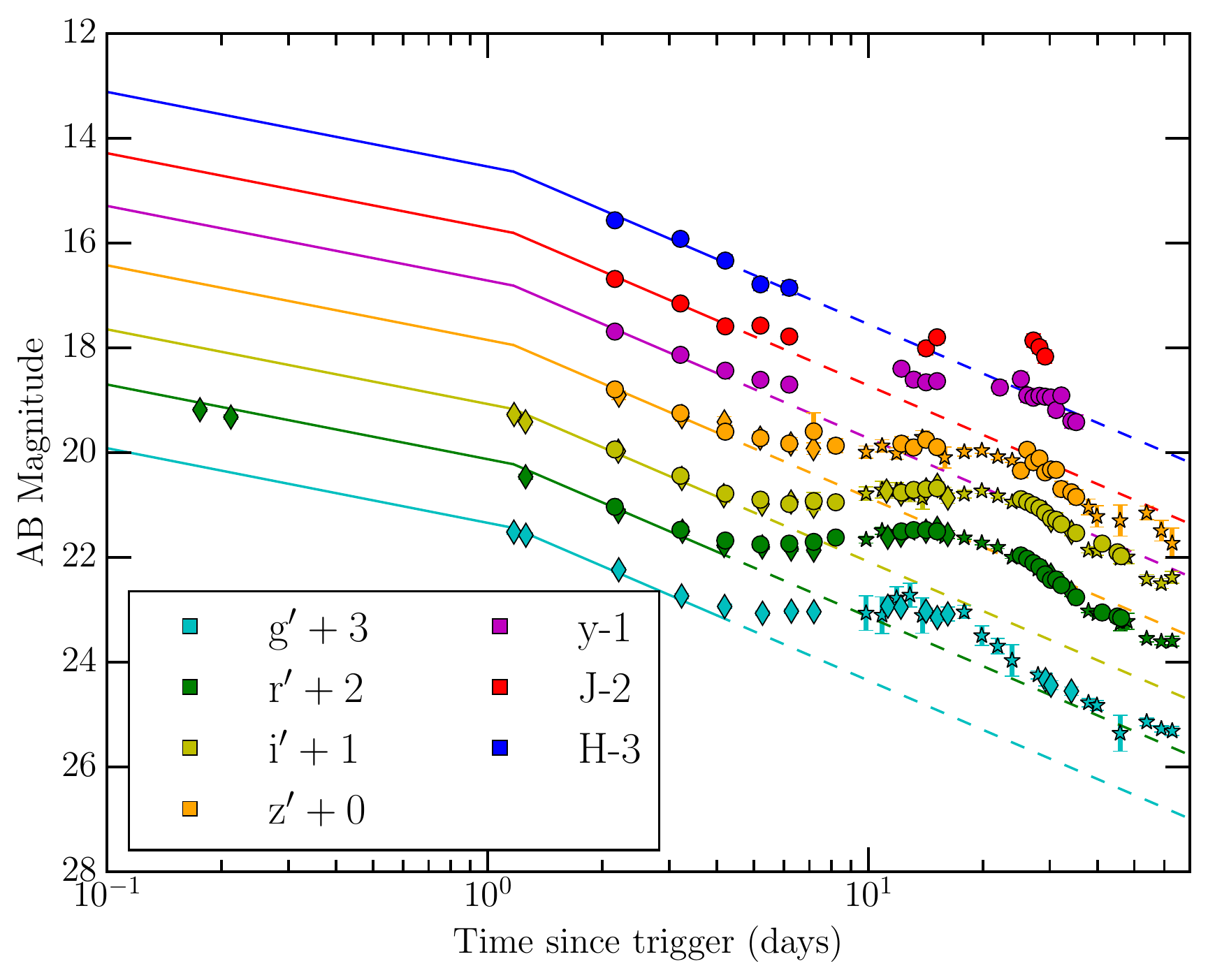}
\caption{ Optical and NIR light curves of GRB\,130702A, corrected for Galactic extinction and host extinction (assuming $A_{V,\mathrm{host}} = 0.10$ mag).  Note that the 
host-galaxy contribution is not removed.  Diamonds are P48/P60 data, circles are RATIR data, and stars are Liverpool data; error bars are overplotted on all datapoints.  Solid lines indicate power-law fits with  $\alpha_1 = 0.57$ and $t_b = 1.17$\,d taken from \citet{Singer:2013a} and $\alpha_2 = 1.25$ from the XRT power-law decay index.  Dashed lines mark the extrapolated power law where we assume $\alpha_2$ for all times beyond $t_b$.   }
\label{fig:lcurve}
\end{figure*}

%%%%%%%%%%%%%%%%%%%%%%%%%%%%%%%%%%%%%%%%%%%%%%

\section{GRB\,130702A/SN 2013dx}

GRB\,130702A was detected by the Gamma-Ray Burst Monitor (GBM; 
\citealt{mlb+09}) on the \textit{Fermi} satellite at 00:05:23.079 
on 2013 July 2 \citep{Collazzi:2013}.  As observed by the GBM, the 
prompt-emission duration\footnote{$T_{90}$ is defined as the 
time over which a burst emits from 5\% of its total measured counts 
to 95\%.} was $T_{90} \approx 59$\,s (50--300\,keV).  High-energy 
emission was also detected by the \textit{Fermi} Large Area Telescope
(LAT; \citealt{aaa+09,Cheung:2013}), as well as by Konus-Wind
\citep{GCN.14986}.  We adopted the Konus-Wind 20--1200\,keV fluence
of $f_{\gamma} = (6.7 \pm 0.8) \times 10^{-6}$\,erg\,cm$^{-2}$ 
(assuming a power-law spectrum with photon index $\Gamma = 1.87 \pm
0.11$).

Employing the wide-field imaging and rapid transient identification
capabilities of the Intermediate Palomar Transient Factory 
(iPTF; \citealt{lkd+09,rkl+09}), \citet{Singer:2013} discovered the
optical afterglow of GRB\,130702A.  The source, also referred to
as iPTF13bxl, is located at (J2000.0) coordinates $\alpha = 
14^\mathrm{h} 29^\mathrm{m} 14^\mathrm{s}.78$, $\delta = 
+15^{\circ} 46\arcmin 26\farcs4$.

Subsequently, the redshift of GRB\,130702A was determined to be
$z = 0.145$ based on the detection of narrow host-galaxy emission lines 
([\ion{O}{3}] and H$\alpha$) at the afterglow location
\citep{Mulchaey:2013,Mulchaey:2013a,DAvanzo:2013}.  Several other 
galaxies at or near this redshift located in the field indicate that  
GRB 130702A occurred in a group or cluster environment, 
which is highly unusual for a GRB  \citep{Kelly:2013,DElia:2015}.
The GRB host galaxy may be a metal-poor satellite of an adjacent 
massive spiral (SDSSJ142914.57+154619.3), which has an offset of 
only $\sim19$\,kpc in projected distance and $<60$\,km\,s$^{-1}$ 
in line-of-sight velocity \citep{Kelly:2013}. 

At $z = 0.145$, the observed Konus-Wind fluence corresponds to
an isotropic energy release of $E_{\gamma,\mathrm{iso}} = 
6.4_{-1.0}^{+1.3} \times 10^{50}$\,erg (1\,keV to 10\,MeV in the rest frame).
  This places GRB\,130702A securely between the low-luminosity class of events 
represented by GRB\,980425 / SN 1998bw and typical cosmologically distant
events with $E_{\gamma,\mathrm{iso}} \gtrsim 10^{52}$\,erg.

\citet{Butler:2013} reported a flattening of the optical afterglow 5.26\,d after the burst.  A spectrum taken $\sim6$\,d after the burst showed broad features resembling those of SN 1998bw \citep{Schulze:2013}. \citet{Cenko:2013} and \citet{DElia:2013} obtained spectra that confirmed the presence of an emerging SN, dubbed SN 2013dx, and identified similarities with SN 1998bw and SN 2006aj.  \citet{DElia:2015} (hereafter D15) reported GRB\,130702A/SN 2013dx light curve, spectra, and SN energetics properties with which we will compare throughout this paper.

\begin{deluxetable*}{rccccccc}
\tabletypesize{\footnotesize}
\tablecolumns{8}
\tablewidth{0pt}
\tablecaption{Log of Spectroscopic Observations}
\tablehead{
\colhead{$\Delta t$} & \colhead{Instrument} & \colhead{Exposure} & 
\colhead{Wavelength Range} & \colhead{Slit} & \colhead{Grating/Grism} & 
\colhead{Airmass} \\
\colhead{(d)} & & \colhead{(s)} & \colhead{(nm)} & \colhead{(\arcsec)} &
& & }
\startdata
1.17 & DBSP & 1800 & 340--1000 & 1.0 & 1200/5000 $+$ 1200/7100 & 1.06 \\
3.25 & DBSP & 1800 & 340--1000 & 1.0 & 600/4000 $+$ 316/7500 & 1.36 \\
6.22 & DBSP & 1800 & 340--890 & 2.0 & 600/4000 $+$ 600/10000 & 1.25 \\
9.33 & DEIMOS & 600 & 450--950 & 1.0 & 600/7500 & 1.19 \\
11.34 & DEIMOS & 600 & 490--1010 & 1.0 & 600/7500 & 1.32 \\
14.21 & DBSP & 1200 & 350--1000 & 1.5 & 600/4000 $+$ 316/7500 & 1.27 \\
31.28 & DEIMOS & 900 & 450--950 & 1.0 & 600/7500 & 1.30 \\
33.27 & LRIS & 1200 & 330--1020 & 1.0 & 400/3400 $+$ 400/8500 & 1.30 \\
330.39 & LRIS & 1460 & 330--1020 & 1.0 & 400/3400 $+$ 400/8500 & 1.04 \\
\label{tab:spec}
\tablecomments{$\Delta$t is the time from \textit{Fermi} trigger in observer frame.}
\enddata
\end{deluxetable*}

%%%%%%%%%%%%%%%%%%%%%%%%%%%%%%%%%%%%%%%%%%%%%%

\section{Observations and Data Reduction}

\subsection{Optical/Near-Infrared Imaging}
We obtained optical and near-infrared (NIR) images of the location of 
GRB\,130702A / SN 2013dx with the robotic Palomar 60\,inch telescope
(P60; \citealt{cfm+06}), the 2\,m Liverpool Telescope (LT; 
\citealt{ssr+04}), the Reionization
and Transients Infrared/Optical camera on the 1.5\,m Harold L.~Johnson
Telescope (RATIR; \citealt{Butler:2012, Watson:2012}), the Large Monolithic 
Imager on the 4.2\,m Discovery Channel Telescope (LMI/DCT), and the
Low Resolution Imaging Spectrometer (LRIS; \citealt{occ+95}) on the 
10\,m Keck-I telescope.  
Additionally, we included early-time Palomar 48-inch $r^\prime$ 
observations (0.17\,d and 0.21\,d after the GRB trigger) from \citet{Singer:2013a}. 
The reduction procedures for each individual facility are described
below, while the resulting photometry is presented in 
Table~\ref{tab:phot} and plotted in Figure~\ref{fig:lcurve}.  
Photometry from different telescopes is calibrated to same stars for uniform 
calibration.  After removing extinction, afterglow, and host galaxy (see \S~\ref{sec:photiso}), the 
cross-calibration errors are $\sim 0.03$--0.05 mag (approximately 3--5\% in flux). 

\subsubsection{P60 Photometry}
P60 observed the location of GRB\,130702A in the $g^{\prime}$, $r^{\prime}$,
$i^{\prime}$, and $z^{\prime}$ filters beginning 1.17\,d after the 
\textit{Fermi} GBM trigger.  Basic CCD reductions are provided in real time
by a custom IRAF\footnote{IRAF is distributed by the National Optical 
Astronomy Observatory, which is operated by the Association of 
Universities for Research in Astronomy (AURA), Inc. under cooperative agreement 
with the National Science Foundation (NSF).}/\texttt{PyRAF}\footnote{See 
\url{http://www.stsci.edu/institute/software_hardware/pyraf}.} pipeline.  At later times ($\Delta t \gtrsim 3$\,d), images were stacked with 
{\tt SWarp} \citep{Bertin:2002} on a nightly basis to increase the signal-to-noise
ratio (SNR).  We performed aperture photometry at the afterglow location,
calibrating with respect to nearby point sources from the Sloan Digital
Sky Survey (SDSS; \citealt{Aihara:2011}).  

\subsubsection{LT Photometry}
LT began observing the location of GRB\,130702A with the IO:O CCD
camera 9.87\,d after the GBM trigger.  Observations were obtained
in the $g^{\prime}$, $r^{\prime}$, $i^{\prime}$, and $z^{\prime}$ 
filters.  Standard reduction techniques were applied to detrend the
data, and photometry was performed in the same manner as for the 
P60 images (including the same SDSS reference stars for photometric
calibration).

\subsubsection{RATIR Photometry}
\label{sec:RATIR}
RATIR obtained simultaneous multi-color 
($r^{\prime}i^{\prime}z^{\prime}yJH$) imaging of the location
of GRB\,130702A beginning 2.16\,d after the GBM trigger. 
The RATIR data were reduced using an automatic \texttt{python} 
pipeline with bias subtraction and twilight-sky flat fielding.  
Given the lack of a cold shutter in RATIR's design, IR dark frames 
were not available. Laboratory testing, however, confirmed that the 
dark current is negligible in both IR detectors \citep{Fox:2012}.
Astrometric solutions were calculated from \texttt{astrometry.net} 
\citep{Lang:2010} and the individual frames are stacked using {\tt SWarp}.

We performed aperture photometry on the resulting stacked images using 
{\tt Sextractor} \citep{Bertin:1996} with an inclusion radius determined 
from the median full width at half-maximum intensity (FWHM) of the images.  
The resulting instrumental magnitudes were 
compared to SDSS in the optical and 2MASS \citep{Skrutskie:2006} in the 
NIR to calculate zeropoints.  For the $y$ band, we created a spectral energy
distribution (SED) from the combination of optical and NIR catalog 
sources and interpolated to the appropriate wavelength.  To place all
photometry on the AB system, we used the $J$- and $H$-band offsets 
from \citet{Blanton:2007}.

\subsubsection{Keck/LRIS Photometry}
\label{sec:lrisphot}
\begin{figure*}[ht]
\centering
\includegraphics[width=5in]{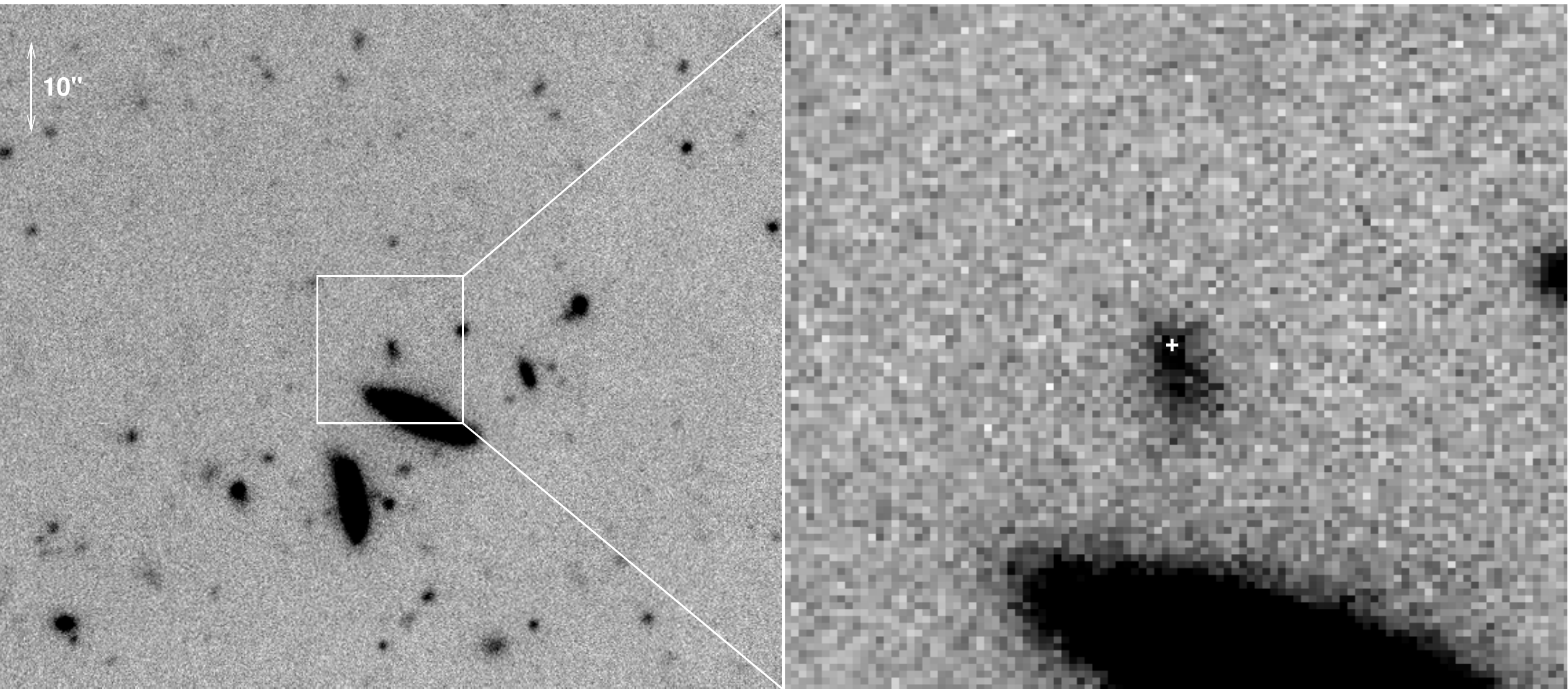}
\caption{Keck/LRIS $g^{\prime}$-band image of the host galaxy of GRB\,130702A,
obtained at $\Delta t = 330.48$\,d after the GBM trigger (i.e., when the
afterglow and SN emission had faded away).  The 
location of the transient is displayed in the inset with the white cross.
The dwarf host is clearly elongated in the N-S direction, with the bulk of the star formation
(as evidenced by the transient location and the nebular emission lines)
apparent in the northern component.  The image is oriented with N up and E 
to the left.}
\label{fig:host}
\end{figure*}

The location of GRB\,130702A was observed with Keck/LRIS on 2014 May 28
($\Delta t = 330$\,d) in the $u^{\prime}$, $g^{\prime}$, and $R$-band
filters.  The resulting images were reduced using the \texttt{LPipe}
package\footnote{See \url{http://www.astro.caltech.edu/~dperley/programs/lpipe.html}
for details.}.  Because the host galaxy was clearly resolved in some
of the better-seeing images (FWHM = 0\farcs75),
we adopted an aperture radius of 1\farcs5 to incorporate all of the flux
from the visible extent of the galaxy (Figure~\ref{fig:host}).  Photometric
calibration was performed relative to point sources from SDSS.

\subsubsection{Keck/MOSFIRE Photometry}
We imaged the location of GRB\,130702A with the Multi-Object Spectrometer
For InfraRed Exploration (MOSFIRE; \citealt{mse+12}) on the 10\,m Keck 
I telescope on 2014 June 16.  Images were obtained in the
$J$ and $K_{s}$ filters and reduced using custom IDL scripts.  We performed
aperture photometry using a 1\farcs5 inclusion radius (see \S~\ref{sec:lrisphot}),
with photometric calibration relative to 2MASS.

\subsubsection{LMI/DCT Photometry}
The location of GRB\,130702A was observed with LMI/DCT on 2015 March 27
($\Delta t = 633$\,d) in the $g^{\prime}$, $r^{\prime}$, $i^{\prime}$, 
and $z^{\prime}$ filters.  The resulting images were detrended with a 
custom IRAF pipeline.  Individual frames were astrometrically aligned 
with {\tt Scamp} \citep{Bertin:2006} and coadded using {\tt SWarp}.  
Photometry was calculated in the manner described in 
\S~\ref{sec:RATIR}.

\begin{figure*}[ht]
\centering
\includegraphics[width=7in]{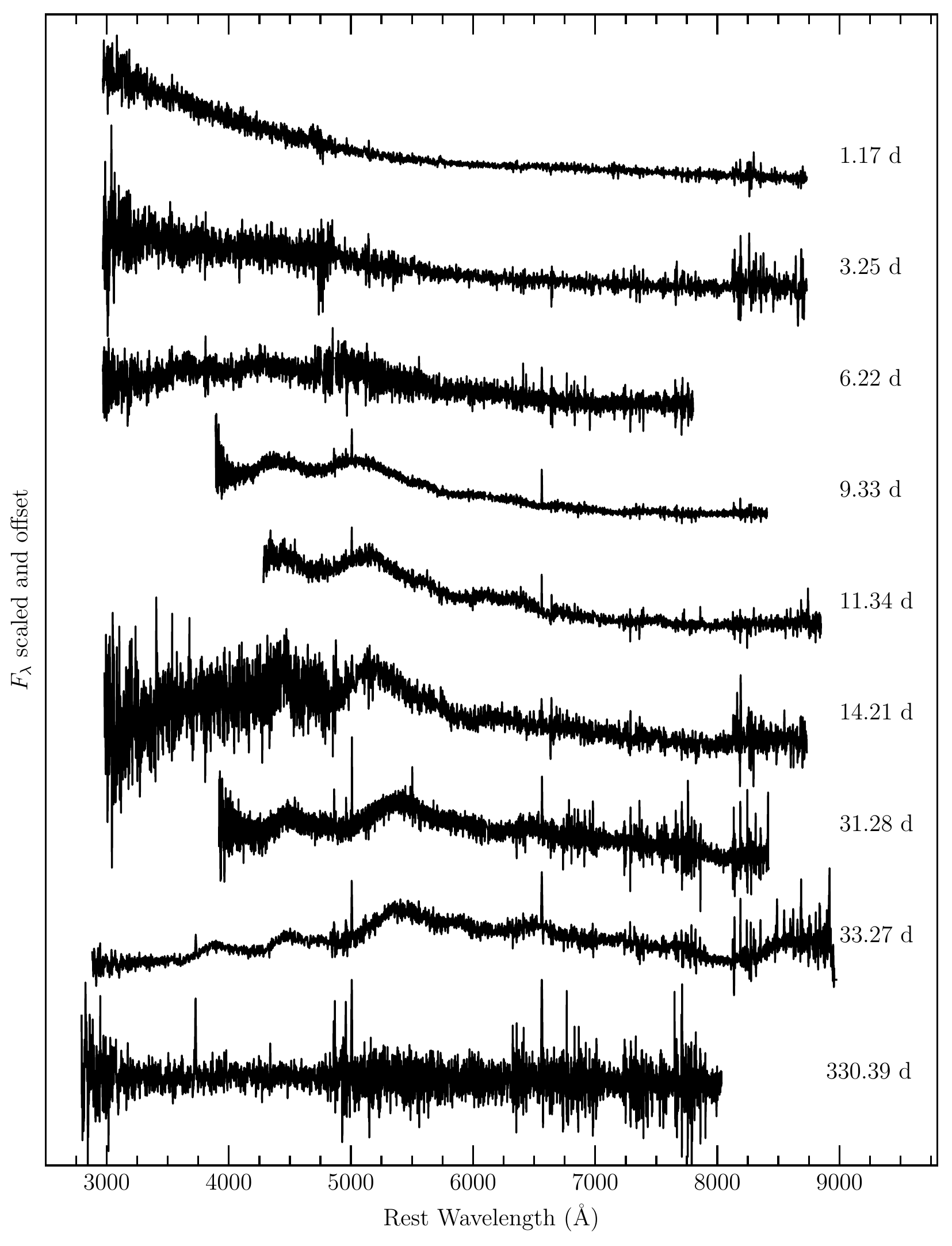}
\caption{Spectra uncorrected for slit losses, extinction, afterglow, or host-galaxy contamination.  Early-time spectra are dominated by the afterglow component.  The broad features
associated with SN 2013dx become visible after about a week.}
\label{fig:spectraraw}
\end{figure*}

\subsection{Optical Spectroscopy}
We obtained a series of optical spectra of GRB\,130702A, beginning at
$\Delta t = 1.17$\,d after the \textit{Fermi}-GBM trigger, with the 
Double Spectrograph (DBSP; \citealt{og82}) on the 5\,m Palomar Hale 
telescope, Keck/LRIS on Keck-I, and the 
DEep Imaging Multi-Object Spectrograph (DEIMOS; \citealt{fpk+03})
on the 10\,m Keck-II telescope.  An observing log is presented
in Table~\ref{tab:spec}.  All spectra were obtained with the slit
oriented at the parallactic angle to minimize differential losses due
to atmospheric dispersion (\citealt{f82}; though note also that LRIS
employs an Atmospheric Dispersion Corrector to further mitigate
against differential slit losses).  The resulting reduced one-dimensional 
spectra are displayed in Figure~\ref{fig:spectraraw}.

\begin{figure}[t]
\centering
\includegraphics[width=3.5in]{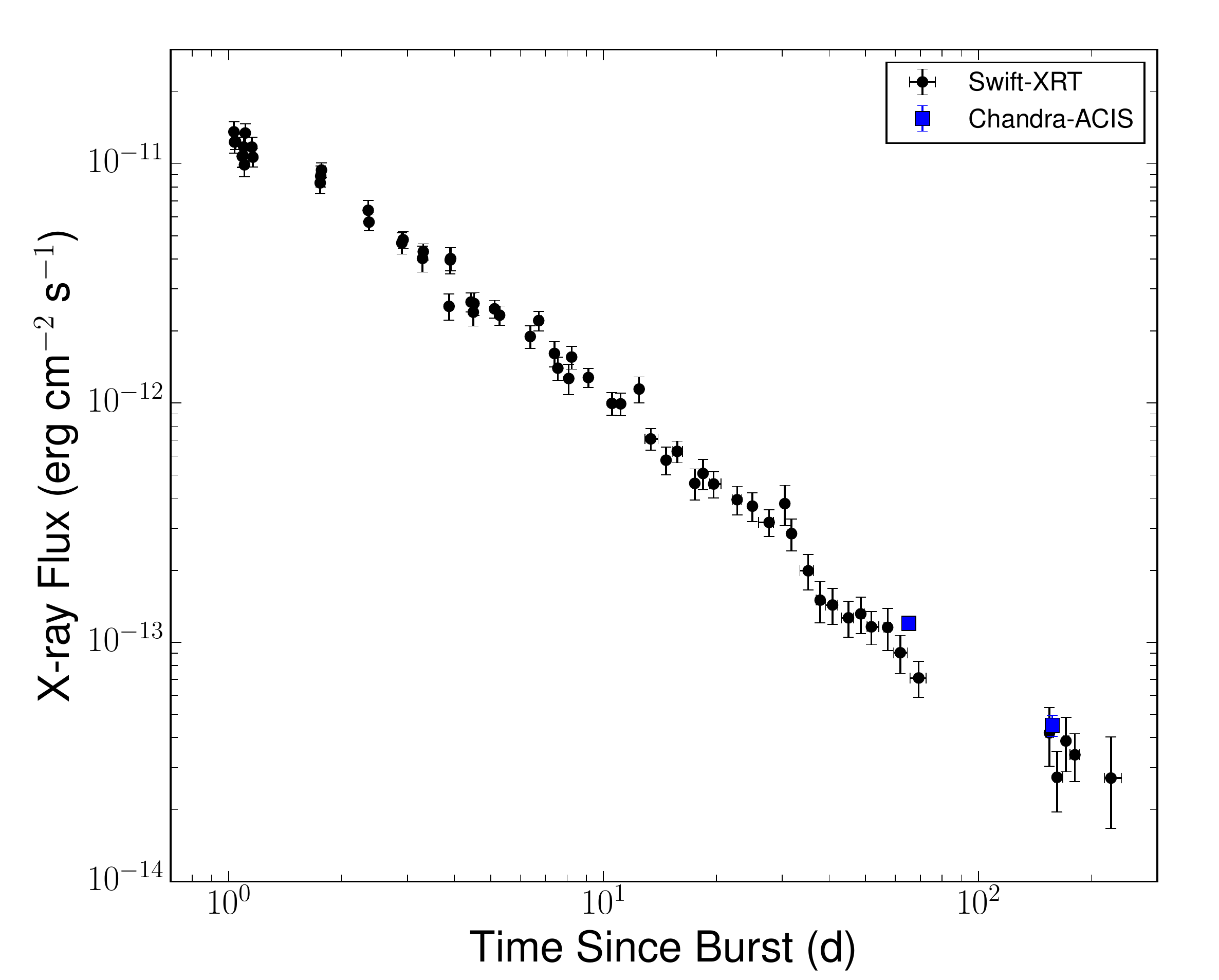}
\caption{X-ray light curve of GRB\,130702A.}
\label{fig:xray_lc}
\end{figure}

\begin{figure*}[ht]
\centering
\includegraphics[width=7in]{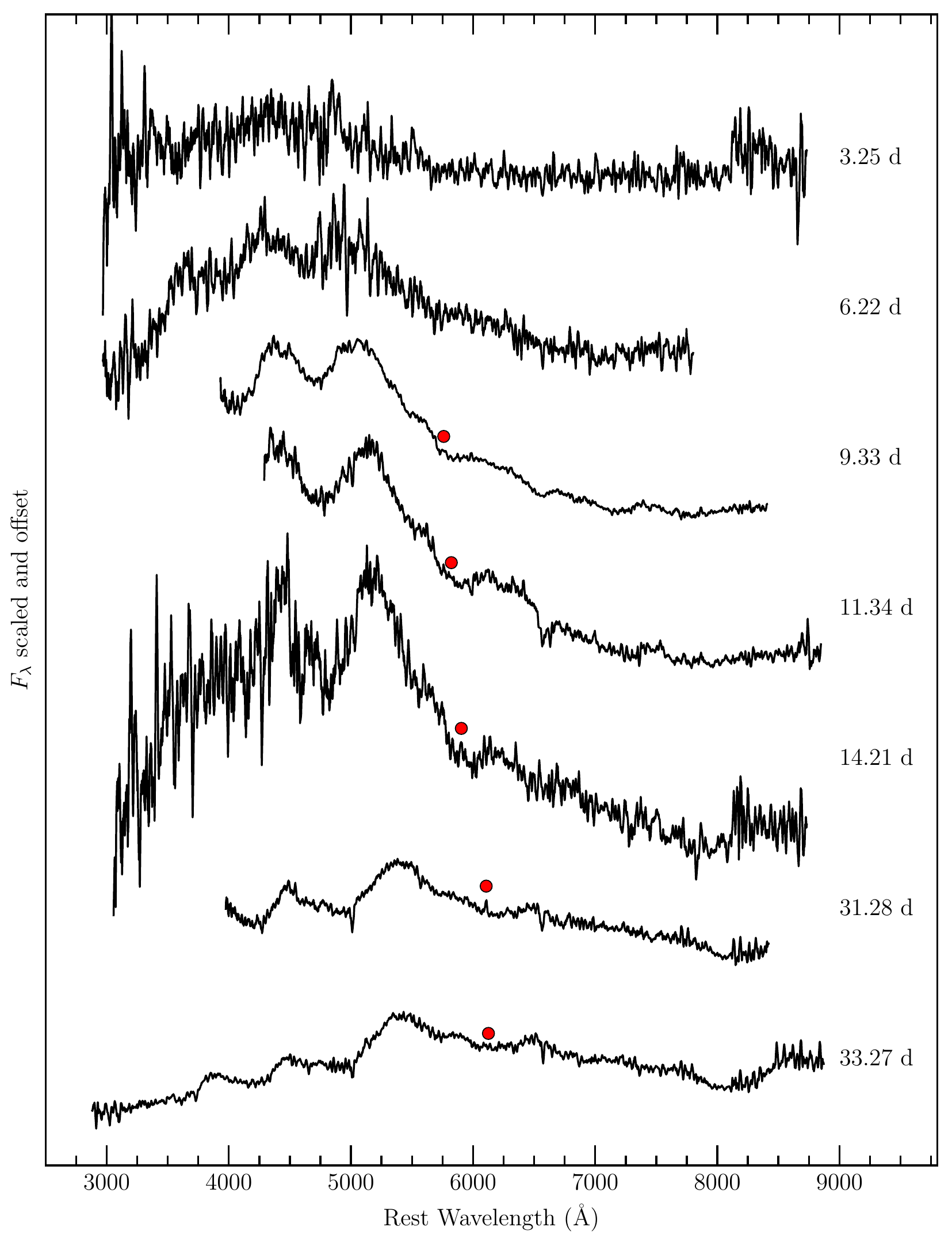}
\caption{SN 2013dx spectra with host galaxy and GRB afterglow removed.  They are smoothed using a Savitzky-Golay filter with a 30\,\AA\ window.  We excluded the spectrum at $t=1.17$\,d because it is dominated by the afterglow. Red filled circles mark the position of \ion{Si}{2} $\lambda$6355 calculated in \S~\ref{sec:vel}, used to determine the photospheric velocity of the ejecta.  The uncertainties are smaller than the symbol width; they are 20\,\AA\ for $t=9.33$\,d and 10\,\AA\ for subsequent epochs. }
\label{fig:spectra}
\end{figure*}

All spectra were reduced using standard routines and optimally 
extracted \citep{h86} within the IRAF environment (see, e.g., 
\citealt{cfp+08} for details).  A dispersion solution was computed 
using calibration spectra of comparison lamps, and then adjusted 
for each individual exposure using night-sky emission lines.  For the LRIS 
and DEIMOS spectra, sky background emission was subtracted using
the algorithm described by \citet{k03}.
Telluric atmospheric absorption features were
removed using the continuum from spectrophotometric standard stars
\citep{wh88,mfh+00}.  Finally, a sensitivity function was applied
using observations of spectrophotometric standards at a
comparable airmass.  We caution that the final Keck/LRIS spectrum 
obtained at $\Delta t = 330.39$ suffered from a failure of the blue 
shutter, which may impact the relative-flux calibration.

Upon publication of this manuscript, all one-dimensional spectra will be made publicly
available via the Weizmann Interactive Supernova data REPository
(WISeREP; \citealt{yg12}).

\subsection{X-Ray Observations}
The afterglow of GRB\,130702A was observed by the X-Ray Telescope 
(XRT; \citealt{Burrows:2005}) onboard the \textit{Swift} satellite 
\citep{gcg+04} beginning at $\Delta t
= 1.03$\,d after the \textit{Fermi}-GBM trigger (e.g., \citealt{Singer:2013a}).
We downloaded the X-ray light curves from the XRT Light Curve 
Repository\footnote{See \url{http://www.swift.ac.uk/xrt_curves} 
and the associated description in \citet{ebp+09}.}.  The time-averaged
spectrum was well described (W-stat $= 299.10$ for 374 degrees of freedom)
by a power-law model, $\Gamma = 1.84 \pm 0.12$, with no 
evidence for $N_{\rm H}$ in excess of the Galactic value 
($N_{\rm H,Gal} = 1.83 \times 10^{20}$\,cm$^{-2}$; \citealt{Kalberla:2005}). 

We initiated deep X-ray follow-up observations of GRB\,130702A with the \textit{Chandra} 
X-ray  Observatory on 2013 September 5, corresponding to $\Delta t=65.2$\,d 
since trigger (PI R. Margutti). The data were reduced with the CIAO software 
package (version 4.6) and corresponding calibration files. Standard ACIS 
data filtering has been applied.  In 14.9\,ks of observations we find clear 
evidence for X-ray emission at the location of GRB\,130702A, with significance 
$>50\sigma$. The spectrum was well modeled by an absorbed power law with 
$\Gamma=1.66\pm0.15$, consistent with the \textit{Swift}-XRT 
time-averaged spectrum.  We found no evidence for an intrinsic absorption 
component, with a $3\sigma$ limit of $N_{\rm H,host}<1.5\times10^{21}\,\rm{cm^{-2}}$.  
Adopting these spectral parameters, the unabsorbed flux is $F_X=(1.20 \pm 0.08)\times 
10^{-13}\,\rm{erg\,s^{-1}cm^{-2}}$ (0.3--10\,keV).

A second epoch of \textit{Chandra} observations was obtained on 2013 December 
6 ($\Delta t=157.5$\,d) with an exposure time of 34.6\,ks. GRB\,130702A was clearly 
detected with significance $>40\sigma$, which allows us to constrain the 
spectral evolution (or lack thereof) of GRB\,130702A at very late times.  Our 
spectral analysis reveals no evidence for spectral evolution. The best-fitting
power-law index was $\Gamma=1.85\pm0.16$, with $N_{\rm H,host}<1.6\times10^{21}\,\rm{cm^{-2}}$ at $3\sigma$ confidence level. 
The corresponding unabsorbed flux is $F_X=(4.5\pm0.45)\times 10^{-14}\,\rm{erg\,s^{-1}cm^{-2}}$ (0.3--10\,keV).

The X-ray light curve of GRB\,130702A, comprising \emph{Swift}-XRT and \emph{Chandra} observations, is presented in Figure~\ref{fig:xray_lc}.

%%%%%%%%%%%%%%%%%%%%%%%%%%%%%%%%%%%%%%%%%%%%%%

\section{Light-Curve Analysis}
\label{sec:lightcurves}

\begin{figure*}[ht]
\centering
\begin{minipage}[b]{0.45\linewidth}
	\includegraphics[width = 1\linewidth]{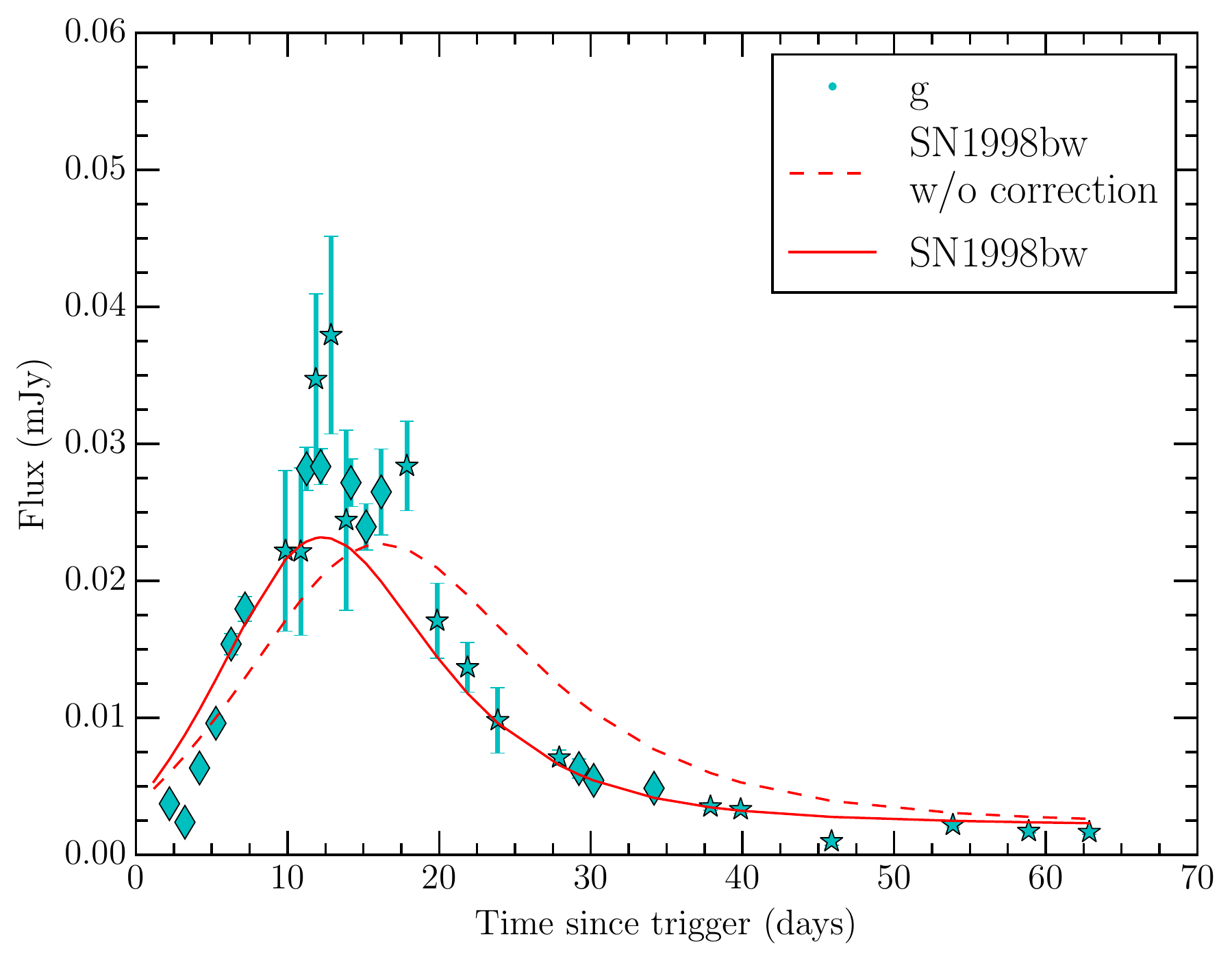}
\end{minipage}
\quad
\begin{minipage}[b]{0.45\linewidth}
	\includegraphics[width = 1\linewidth]{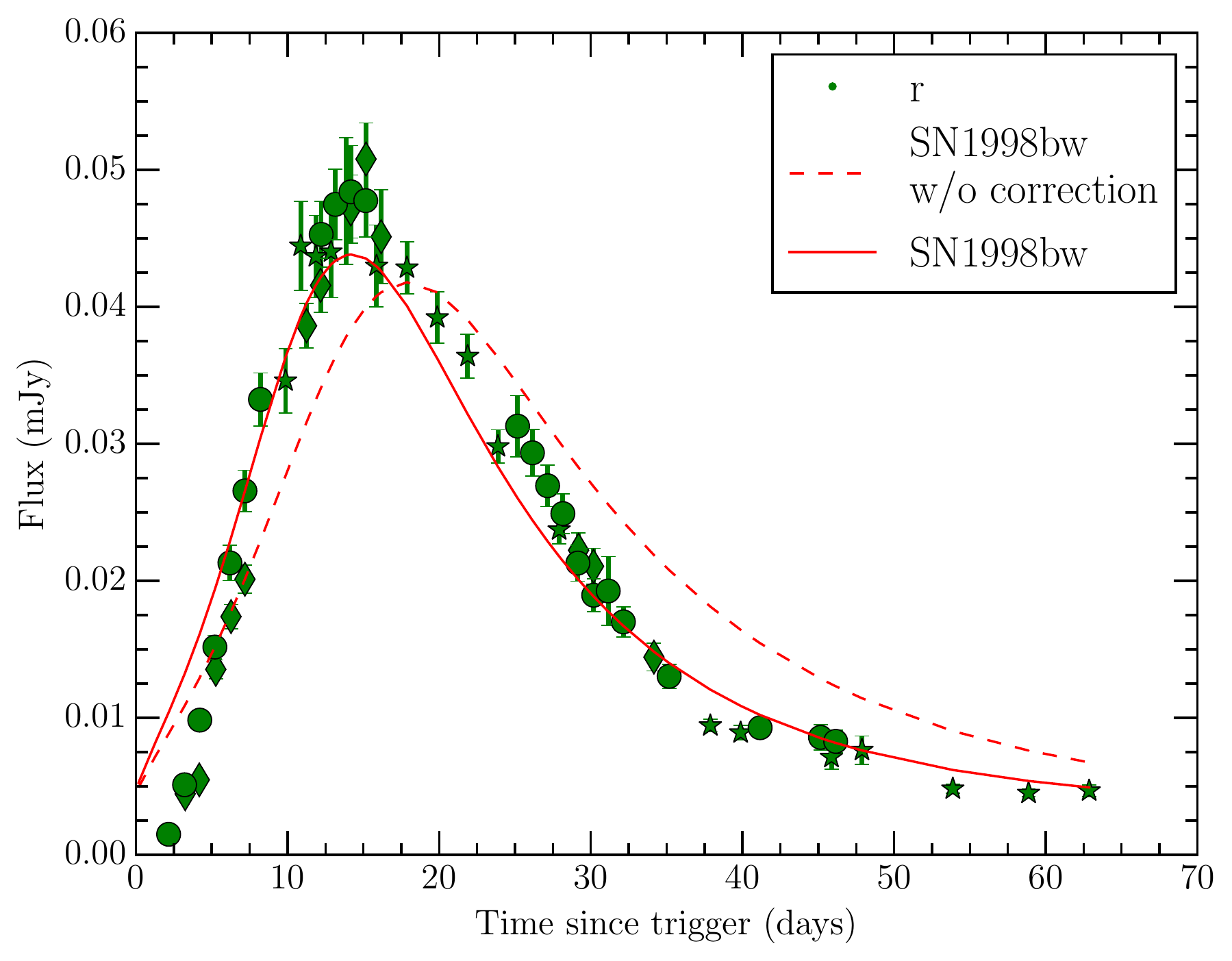}
\end{minipage} 

\begin{minipage}[b]{0.45\linewidth}
	\includegraphics[width = 1\linewidth]{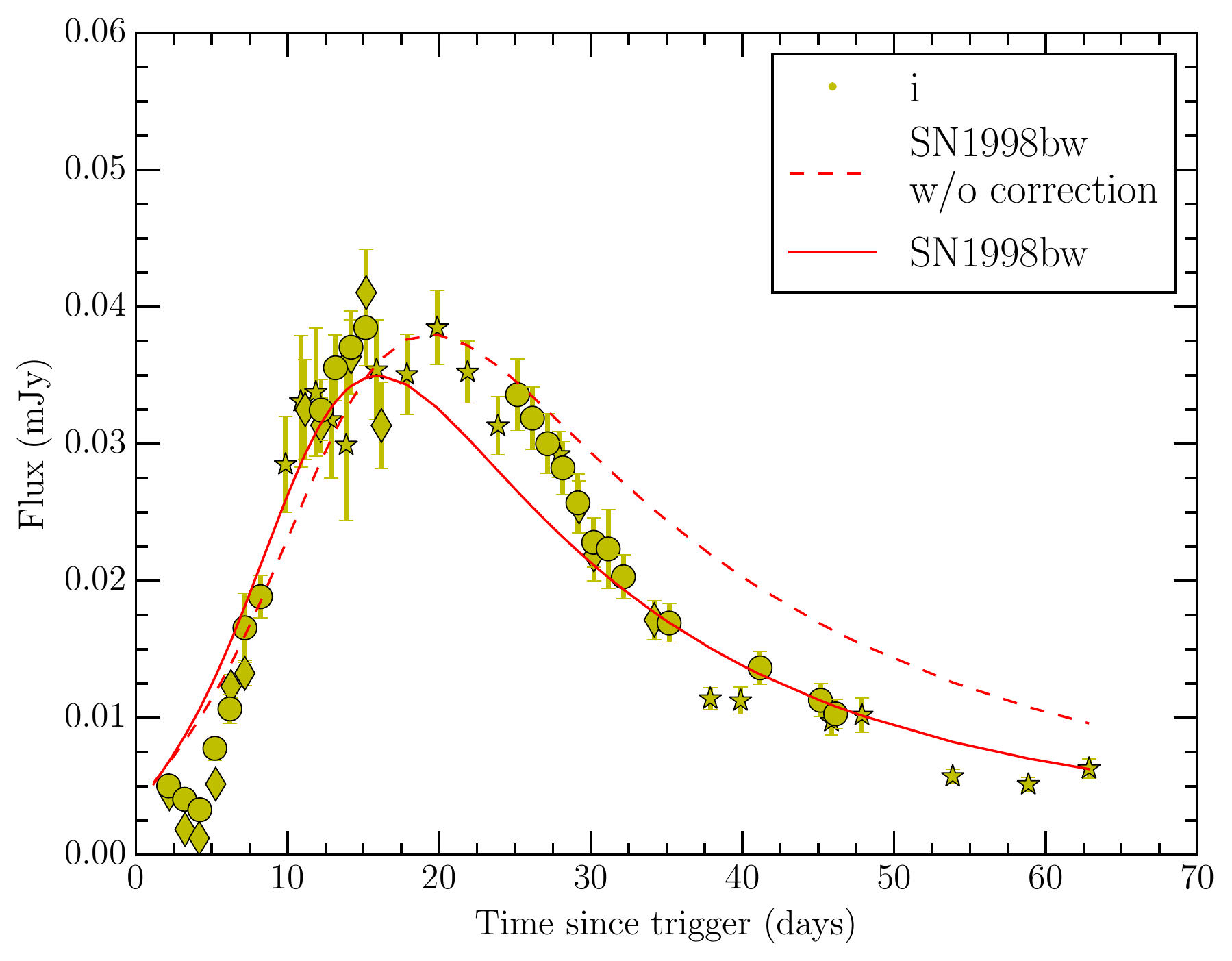}

\end{minipage}
\quad
\begin{minipage}[b]{0.45\linewidth}
	\includegraphics[width = 1\linewidth]{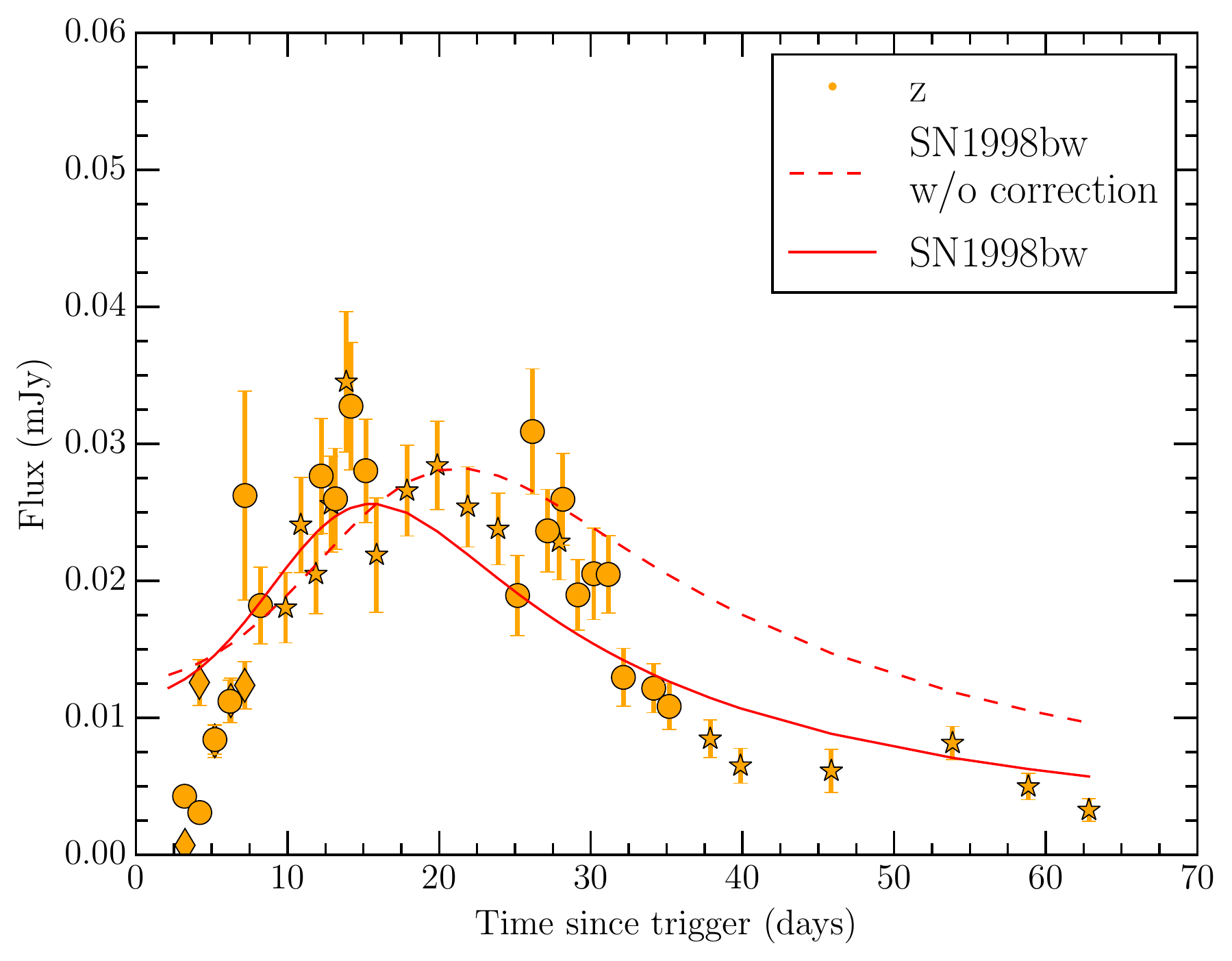}
\end{minipage}
\caption{Observed SN 2013dx $g^{\prime}r^{\prime}i^{\prime}z^{\prime}$ data (top left, top right, bottom left, and bottom right, respectively), with synthetic SN 1998bw light-curve fit (dashed red line) and an optimized synthetic SN 1998bw light-curve fit scaled by a peak-amplitude factor $k$ and time-stretch factor $s$ (solid red line; see \S \ref{sec:sn98bw}) overlaid.  SN 1998bw template created from \citet{Clocchiatti:2011}, \citet{Galama:1998}, \citet{Sollerman:2002}, and \citet{Patat:2001}. Diamonds are P48/P60 data, circles are RATIR data, and stars are Liverpool data.}
\label{fig:sk}
\end{figure*}

\subsection{Isolating the Supernova Component}
\label{sec:photiso}
Emission from the location of GRB\,130702A results from three distinct components:
the GRB afterglow, the associated SN, and the underlying host galaxy.
Here we try to isolate the emission resulting from the associated SN,
including a proper accounting for line-of-sight extinction,
in order to study the properties of SN 2013dx.

First, we correct our broadband photometry for extinction, 
both in the Milky Way and in the host galaxy.
For the Galactic component, we employ the dust-map calibration of 
\citet{Schlafly:2011}, resulting in $E(B-V)_{\mathrm{MW}} = 0.038$\,mag,
and the Milky Way extinction law of \citet{ccm89}.  In order to estimate the 
host extinction, $A_{V,\mathrm{host}}$, we create an SED at $\Delta t =2.25$\,d from linear interpolation.  We assume 
the observed emission at this stage will be dominated by the
(synchrotron) afterglow, and thus we fit the SED to a simple power-law
model of the form $f_{\nu} \propto \nu^{-\beta}$ (e.g., \citealt{spn+98}).  
We incorporate $A_{V,\mathrm{host}}$ as a free parameter, assuming a 
Small Magellanic Cloud (SMC)-like extinction law \citep{Pei:1992}.  We find 
$A_V = 0.13 \pm 0.23$\,mag and $\beta = 0.52 \pm 0.19$ with a reduced 
$\chi^{2}_{\mathrm{red}} = 0.83$.  Adopting Large Magellanic Cloud (LMC) and 
Milky Way (MW) dust extinction laws did not alter the derived parameters
or fit quality.  For the rest of this paper we assume $A_{V,\mathrm{host}} 
= 0.10$\,mag.  This is consistent with other GRB host-extinction values: 
50\% of GRBs have $A_{V,\mathrm{host}} < 0.4$\,mag and 87\% of GRBs have
$A_{V,\mathrm{host}} < 2$\,mag \citep{Covino:2013}.

Next, we attempt to remove any contribution from the afterglow.  \citet{Singer:2013a}
modeled the early-time optical emission ($\Delta t \lesssim 4$\,d) as a broken
power law with an initial decay index of $\alpha_{1} = 0.57 \pm 0.03$
up to the break time, $t_{\mathrm{b}} = 1.17 \pm 0.9$\,d,
after which the model followed a power-law decay index of $\alpha_{2} = 1.05 \pm 0.03$.  
We repeat this analysis with our larger
photometric dataset and find mostly consistent results.  However, even as early as
several days post-trigger, the observed emission will likely have some
contribution from the emerging SN (e.g., the broad features becoming apparent
in the $\Delta t = 3.25$\,d DBSP spectrum in Figure~\ref{fig:spectra}). 
Consequently, the true afterglow decay index may be steeper than indicated here.

Instead, we consider the decay of the corresponding X-ray emission, which 
is unlikely to be contaminated by the SN at $\Delta t \gtrsim 1$\,d.  We fit
the X-ray light curve to a power-law model and find $\alpha_{X} = 1.25 \pm
0.03$.  Combining this with the measured X-ray spectral index from the 
\textit{Swift}-XRT data, $\beta_{X} = 0.84 \pm 0.12$, we can use standard
afterglow closure relations (e.g., \citealt{rlb+09}, and references therein)
to evaluate where the X-rays fall on the broadband synchrotron spectrum. 
The best fit is found for a constant-density circumburst medium with the
X-rays falling below the synchrotron cooling frequency, $\nu_{c}$.  

As a result, the optical bandpass must fall below $\nu_{c}$ as well, which
is consistent with the measured X-ray to optical spectral index of $\beta_{OX}
\approx 0.7$ at $\Delta t = 2$\,d.  Thus, since the optical emission falls on the
same segment of the SED as the X-rays, it should
decay with the same power-law index, $\alpha_{O} = 1.25$.  We further assume
that both the optical afterglow spectral and temporal indices remain 
constant in time over the course of our observations.  We use this model to 
calculate the afterglow contribution for all our photometric observations.  

We note that this is a significantly
shallower decay index than the $\alpha_{O} = 2.2$ adopted by D15, who did 
not incorporate multi-wavelength observations into their afterglow analysis.
A steeper $\alpha_{O}$ may overestimate the SN flux at 
early times, but is negligible at the peak and late times when the SN 
is significantly brighter than the afterglow.  

Finally, we must remove the contribution from the underlying host galaxy.
In our best-seeing images at late times (FWHM $\approx$ 0\farcs75), the host
is clearly resolved, with the afterglow/SN location falling on a blue ``knot''
to the north (Figure~\ref{fig:host}).  This location is also responsible
for the nebular emission lines seen in the final Keck/LRIS spectrum
(Figure~\ref{fig:spectraraw}).  

To remove the host contribution from our photometry, we adopt host flux 
values from our late-time DCT ($g^{\prime}r^{\prime}i^{\prime}z^{\prime}$)
and Keck/MOSFIRE ($J$) imaging and directly subtract these from the measured
transient fluxes.  While this does not account for the resolved nature of
the host, since the typical seeing in our SN data ($\sim$ 1\farcs5) 
is comparable to the size of the extended emission, this should have minimal
impact on the resulting photometry.   We do 
not have host-galaxy detections in the \textit{y} and \textit{H} bands, but the 
host contribution is negligible compared to the afterglow and SN components.

The resulting SN light curves are displayed in Figure~\ref{fig:sk}.  The peak 
times of the light curves are useful for constraining theoretical models, in 
particular the convolution of total ejected mass, kinetic 
energy, and opacity of the SN explosion (via the diffusion time).
We measure the rest-frame peak times by fitting a second-order polynomial at 
$7 \leq \Delta t <20$\,d and find the following:
$t_{\mathrm{p}}(g^{\prime}) = 11.7 \pm 0.3$\,d, 
$t_{\mathrm{p}}(r^{\prime}) = 13.2 \pm 0.3$\,d, 
$t_{\mathrm{p}}(i^{\prime}) = 14.7 \pm 0.6$\,d, and 
$t_{\mathrm{p}}(z^{\prime}) = 15.1 \pm 1.6$\,d (statistical uncertainties
only). These values generally agree well with those reported by D15. 
We do not include peak times for $yJH$ because the data are not well
sampled close to the peak.

There is marginal evidence in the $i^{\prime}$-band light curve (Figure~\ref{fig:sk}) 
of a decline in flux at early times ($\Delta t \lesssim 3$\,d).  This is consistent 
with early signatures of shock cooling (e.g., SN 2006aj, \citealt{Campana:2006}; 
SN 2010bh, \citealt{Cano:2011}).  However, shock breakout should be significantly 
stronger in bluer bands and we see no indication of it in either the $g^{\prime}$
or $r^{\prime}$ bands.  The relatively bright optical afterglow of GRB\,130702A,
compared to the optical afterglows of (for example) GRB\,060218 and GRB\,100316D, greatly
complicates isolating the SN component at early times.  Thus, it is difficult
to reach firm conclusions regarding the presence or absence of a shock-breakout
signature.

\begin{deluxetable}{cccc}
\tabletypesize{\footnotesize}
\tablecolumns{4}
\tablewidth{0pt}
\tablecaption{SN 1998bw Template Fits}
\label{tab:sandk}
\tablehead{
\colhead{Filter} & \colhead{$s$} & \colhead{$k$} & \colhead{$\chi_{\rm red}^2$}  }
\startdata

$g^{\prime}$	&  0.76 $\pm$ 0.05 & 1.02 $\pm$ 0.06 & 7.78\\
$r^{\prime}$	&  0.79 $\pm$ 0.01 & 1.05 $\pm$ 0.02 & 4.67\\
$i^{\prime}$	&  0.82 $\pm$ 0.02 & 0.92 $\pm$ 0.03 & 4.07\\
$z^{\prime}$	&  0.74 $\pm$ 0.04 & 0.91 $\pm$ 0.05 & 2.85

\enddata
\label{tab:sk}
\end{deluxetable}

\subsection{Comparison with SN 1998bw}
\label{sec:sn98bw}
Following past studies of GRB-associated SNe in the literature, we next attempt to 
compare SN 2013dx to the well-studied SN 1998bw (associated with GRB\,980425).
We create K-corrected synthetic SN 1998bw light curves in the 
$g^{\prime}r^{\prime}i^{\prime}z^{\prime}$ filters at the redshift of SN 2013dx, 
$z=0.145$, using methods described by \citet{Hogg:2002}.  We utilize SN 1998bw 
photometry and spectra from \citet{Clocchiatti:2011}, \citet{Galama:1998},
\citet{Sollerman:2002},
and \citet{Patat:2001}.  The K-corrected synthetic SN 1998bw light curves were 
also time dilated to match the observer frame of SN 2013dx.  

Owing to gaps in the temporal coverage of SN 1998bw photometry, 
especially in the rising phase, we fit the synthetic SN 1998bw light curve in
each filter with the empirical functional form from \citet{Cano:2011},
\begin{equation}\label{eq:lightcurve}
U(t) = A +pt \bigg( \frac{e^{(-t^{\alpha_1}/F)}}{1+e^{(p-t/R)}} \bigg) + t^{\alpha_2}\textrm{log}(t^{-\alpha_3}),
\end{equation}
allowing $A, p, F, R, p, \alpha_1, \alpha_2, \textrm{and } 
\alpha_3$ to vary.  The resulting SN 1998bw synthetic light curves are plotted
as dashed lines in Figure~\ref{fig:sk}.  

We then assume the light curves of SN 2013dx in each of our four filters
can be modeled by simply varying the peak amplitude ($k$) and stretch
factor ($s$):
\begin{equation}\label{eq:lightcurve2}
L_{\rm 13dx}(t) = k \, U_{\rm 98bw}(t/s).
\end{equation}
The resulting fits are plotted in Figure~\ref{fig:sk}, while the measured
stretch and amplitude values are displayed in Table~\ref{tab:sk}.  We find
that SN 2013dx has a peak flux comparable to that of SN 1998bw in all four
filters reported here (slightly more luminous in the bluer filters, and 
slightly less luminous in the redder filters).  With a constant stretch
value of $s \approx 0.8$ in all four filters, the evolution of SN 2013dx
(in particular the rise time) is noticeably faster than that of SN 1998bw.

However, it is also clear from the fits that SN 1998bw is not an ideal match to
SN 2013dx, especially in the redder $i$- and $z$-band filters.  Given these
important differences, we refrain from drawing any physical inferences
(e.g., $M_{\mathrm{Ni}}$, $E_{\mathrm{K}}$) from these fits, and instead
use the more model-independent bolometric light curve in \S~\ref{sec:SNparams}.

\begin{figure*}[ht]
\centering
\includegraphics[width=7in]{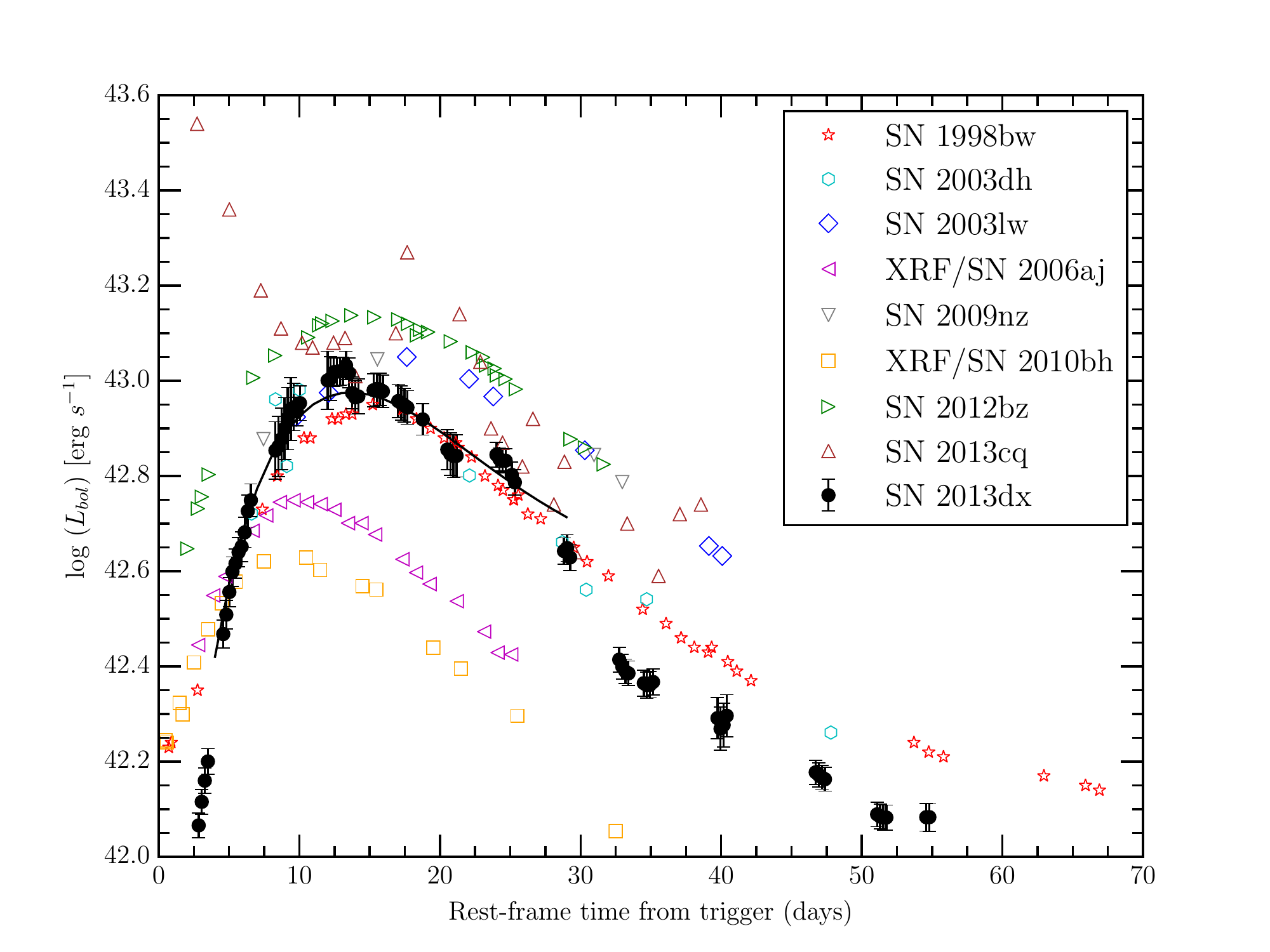}
\caption{$g^{\prime}r^{\prime}i^{\prime}z^{\prime}yJ$ bolometric light-curve fit with Arnett-Valenti relation fit for $4 \leq t \leq 30$\,d (solid black line).  Bolometric light curves of other spectroscopically confirmed GRB-SNe are included for comparison.  Note that these ``bolometric'' light curves are actually ``quasi-bolometric'' and cover different wavelength ranges.}
\label{fig:bolometric}
\end{figure*}

\citet{Cano:2014} and \citet{lbj+14} suggested that GRB-SNe can be used as standardizable candles.  \citet{Cano:2014}  proposed that $s$ and $k$ (measured relative to SN 1998bw) are analogous to the absolute peak SN magnitude and the amount the light curve fades from maximum light to 15\,d later ($\Delta m_{15}$) used in the Phillips relation \citep{Phillips:1993} for SNe~Ia.  We add the uncertainties from our $s$ and $k$ fit in quadrature and find that our measurements of SN 2013dx deviate by $3.7\sigma$ from the \citet{Cano:2014} fits.  The worst fits to \citet{Cano:2014} are for $g^{\prime}$ and $r^{\prime}$ at $3.7\sigma$; however, $z^{\prime}$ is within $0.9\sigma$.  This
further supports the notion that SN 1998bw is not a perfect match for all
GRB-associated SNe.

\subsection{Bolometric Light Curve}
\label{sec:bolometric}

We construct the quasi-bolometric light curve of SN 2013dx using our 
photometry in the $g^{\prime}$, $r^{\prime}$, $i^{\prime}$, $z^{\prime}$,
$y$, and $J$ filters ($H$ had only upper limits at $\Delta t \gtrsim 
5$\,d).  We include synthetic photometry for $\Delta t=31.28 \text{ and } 
33.27$\,d from our spectra (see \S~\ref{sec:speciso}) to supplement photometric 
coverage at these epochs.  We assume a 10\% flux error on all synthetic 
photometry data points.
We convert the extinction, host-galaxy, and afterglow-corrected 
magnitudes to monochromatic fluxes and create SEDs from linear interpolation 
of the data for each epoch between 1\,d and 70\,d with 0.25\,d spacing.  
Epochs that are $> 0.5$\,d from observations are removed to 
mitigate linear-interpolation errors, which only affects observations in 
the $y$ and $J$ bands (see below).  We assume that the flux is 
constant across the bandwidth of each filter and use trapezoidal 
integration to calculate the quasi-bolometric luminosity.  We note that 
our photometry provides coverage over the observer-frame bandpass
0.4--1.35\,$\mu$m (rest-frame 0.35--1.18\,$\mu$m).

At $\Delta t \approx 15$--25\,d and $\Delta t > 35$\,d, the $y$- and $J$-band 
data are relatively sparse.  We therefore calculate the bolometric 
luminosity for the entire light curve both
including $y$ and $J$ ($g^{\prime}r^{\prime}i^{\prime}z^{\prime}yJ$)
and excluding them ($g^{\prime}r^{\prime}i^{\prime}z^{\prime}$) to determine
the NIR to integrated flux ratio.  We find that the fraction of flux at these wavelengths 
increases monotonically as a function of time, from 13\% at $\Delta t \approx
6$\,d to 23\% at $\Delta t \approx 29$\,d.  For epochs when only 
$g^{\prime}r^{\prime}i^{\prime}z^{\prime}$ observations were available, 
we add a fractional NIR contribution for the $y$- and $J$-band from our linear fit.  
At late times we adopt the last NIR ratio measurement at $\Delta t \approx 
29$\,d of $\sim23$\% instead of extrapolating our linear fit. This may 
underestimate the NIR contribution at late times.  Our NIR contribution 
measurements are consistent with an analogous measurement for SN 2008D
\citep{Modjaz:2009}, though slightly smaller than for SN 2009bb and SN 2010bh
\citep{Cano:2011}, which have  
maximum NIR contributions of 35--45\%  at $\Delta t \approx 25$\,d.

The resulting bolometric light curve is displayed in Figure \ref{fig:bolometric}.
Our associated uncertainty measurements incorporate errors from the flux measurements,
as well as bandpass uncertainties, but do not include errors introduced from linear
interpolation.

\begin{figure*}[ht]
\centering
\begin{minipage}[b]{0.45\linewidth}
	\includegraphics[width=1\linewidth]{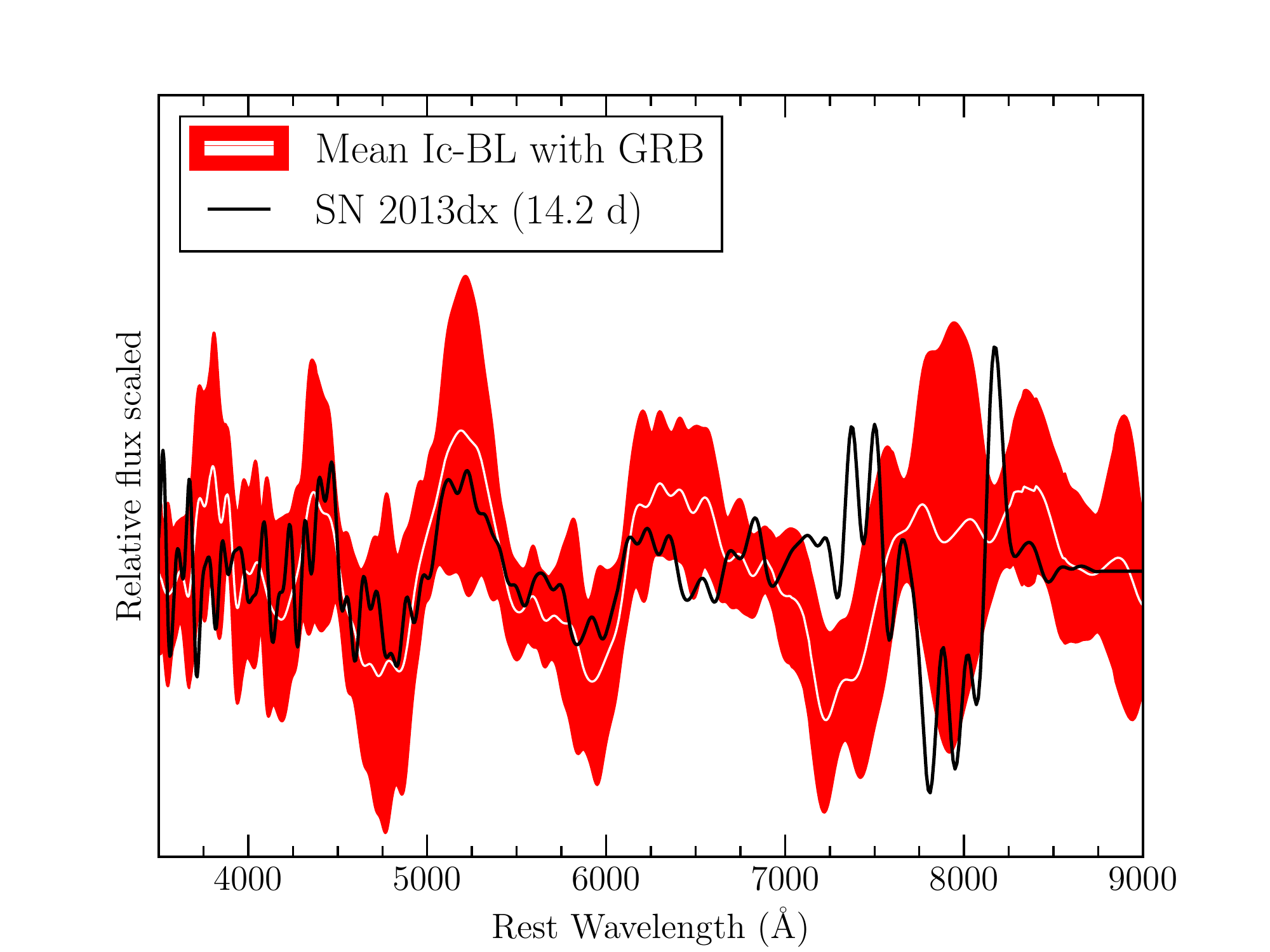}
\end{minipage}
\quad
\begin{minipage}[b]{0.45\linewidth}
	\includegraphics[width = 1\linewidth]{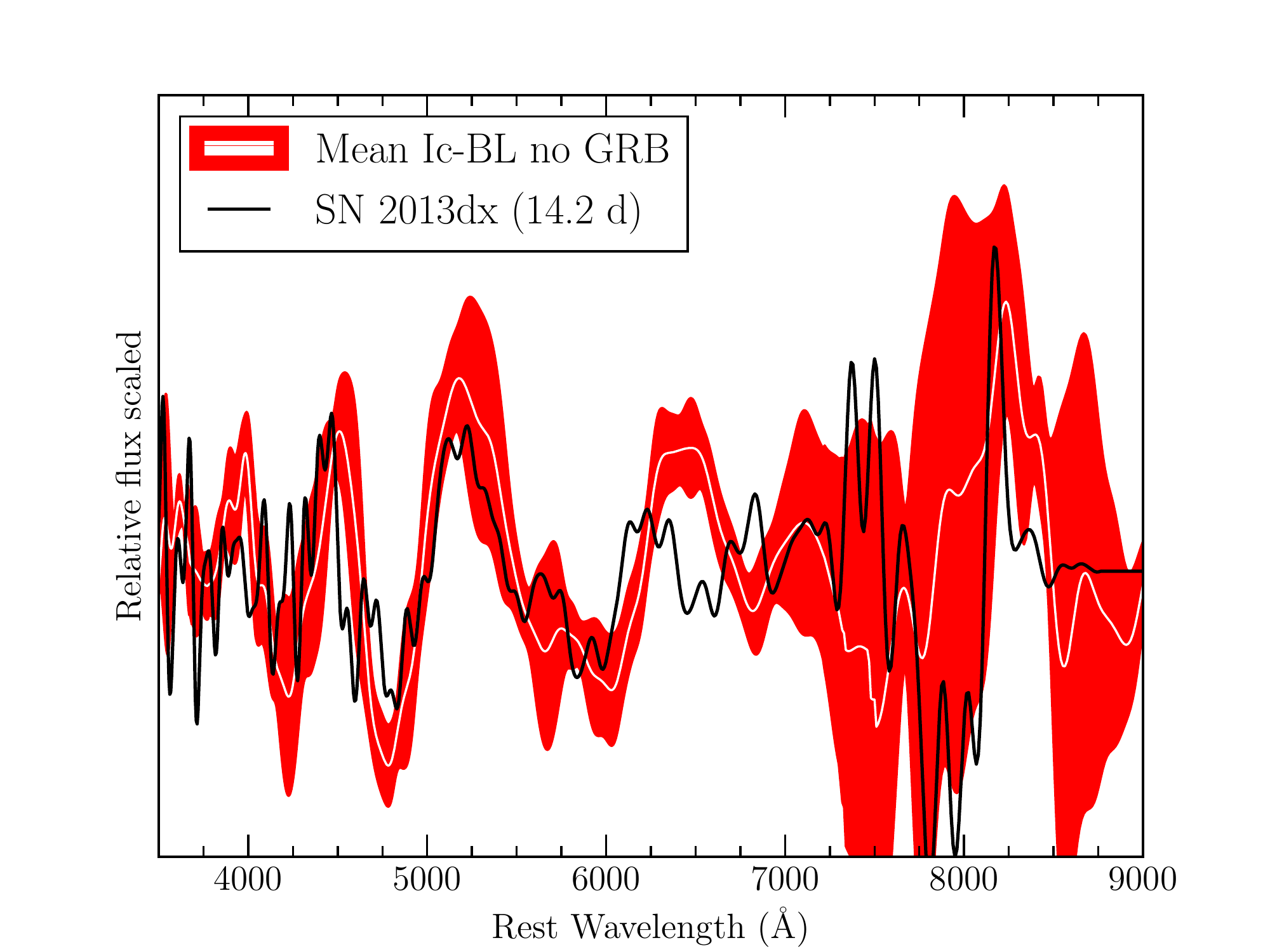}
\end{minipage} 

\begin{minipage}[b]{0.45\linewidth}
	\includegraphics[width=1\linewidth]{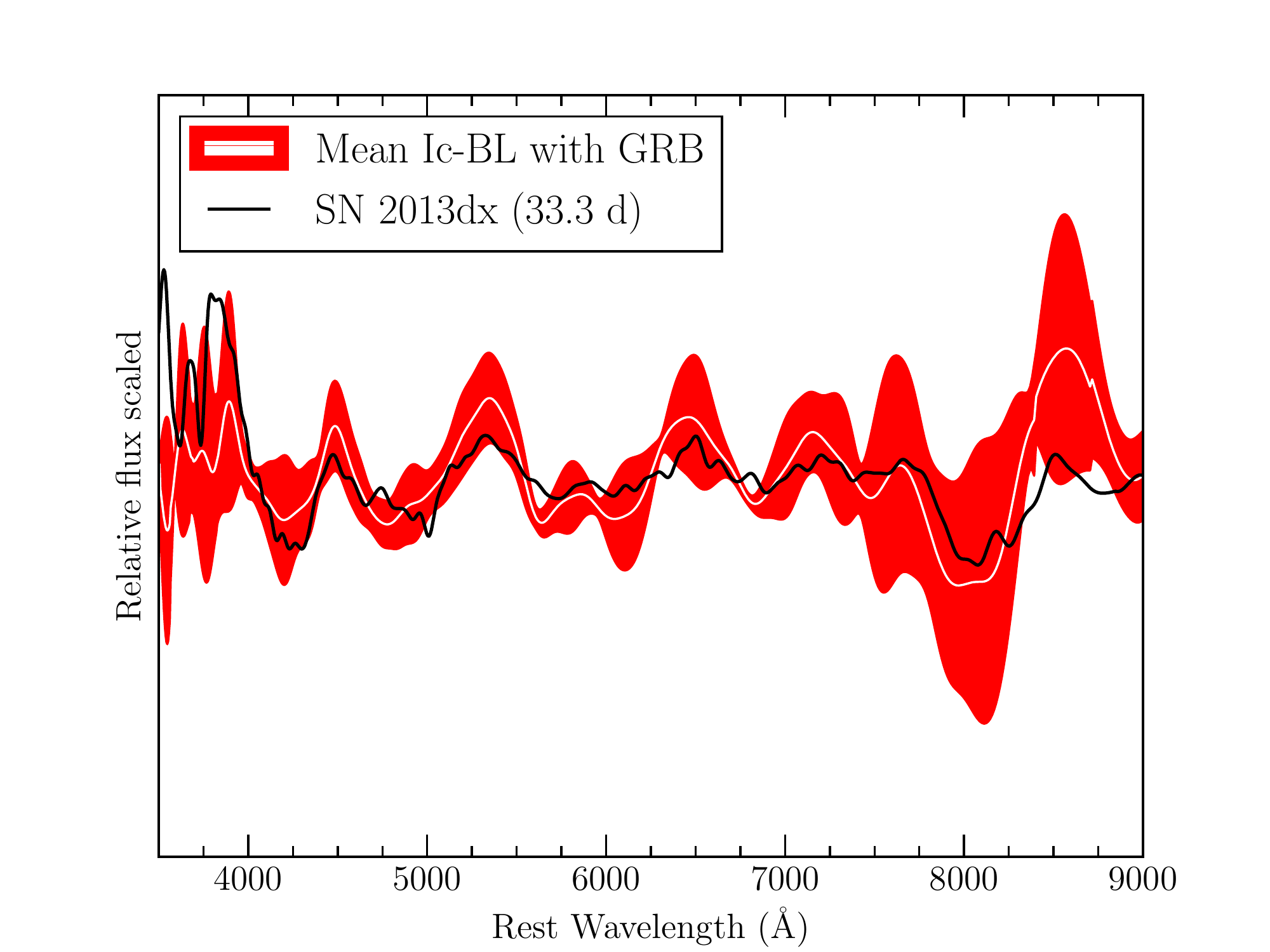}
\end{minipage}
\quad
\begin{minipage}[b]{0.45\linewidth}
	\includegraphics[width = 1\linewidth]{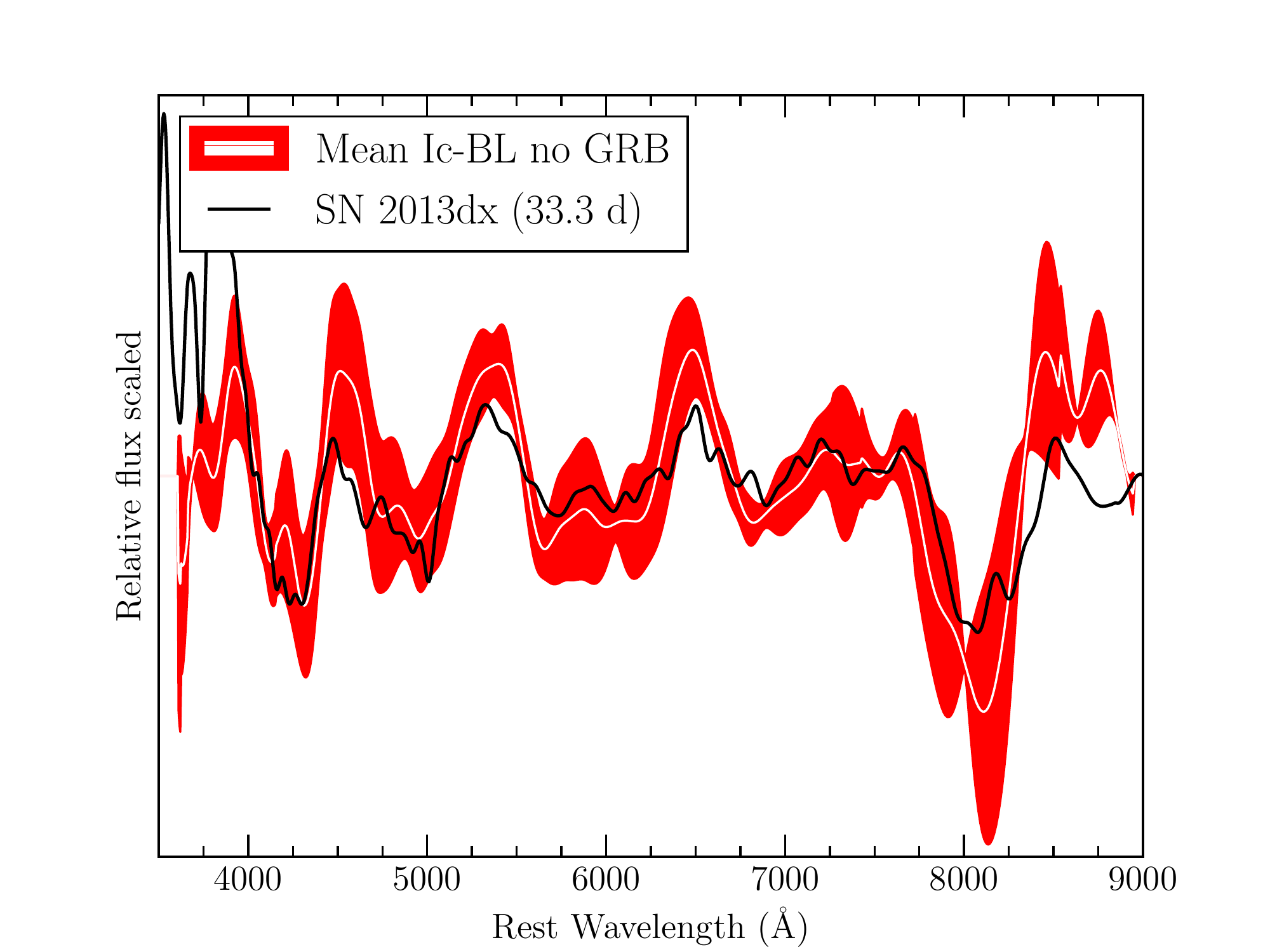}
\end{minipage} 
\caption{SN 2013dx spectrum with pseudocontinuum removed and bandpass filtered, binned to a common logarithmic wavelength scale (black).  The mean (white) and standard deviation (red) of Type Ic-BL spectra from \citet{Modjaz:2015} with (left) and without (right) GRB within 2 rest-frame days of the SN 2013dx spectrum. \textit{(top)} Spectrum taken at $\Delta  t=14.2$\,d, closest to maximum light in $r^{\prime}$ ($\Delta t = 13.2$\,d) and $i^{\prime}$ ($\Delta t=14.7$\,d).   \textit{(bottom)} Spectrum taken at $\Delta  t=33.3$\,d.}
\label{fig:spectramax} 
\end{figure*}

\citet{Lyman:2014} have created a model for  core-collapse SN bolometric corrections 
using two filters for nearby events.  The corrections include ultraviolet and NIR contributions.  
Since SN 2013dx is at $z=0.145$, we use our K-corrected spectra (see \S \ref{sec:speciso})
to extract $g^{\prime}$ and $r^{\prime}$ synthetic photometry.  We apply the methodology
described by \citet{Lyman:2014} and find that this model leads to excellent agreement with our brute-force 
$g^{\prime}r^{\prime}i^{\prime}z^{\prime}yJ$ bolometric light curve.  This confirms that we are not underestimating the ultraviolet and 
NIR contributions in our quasi-bolometric light curve.

%%%%%%%%%%%%%%%%%%%%%%%%%%%%%%%%%%%%%%%%%%%%%%

\section{Spectral Analysis}
\label{sec:spectralanalysis}

\subsection{Isolating the SN Component}
\label{sec:speciso}
In a similar manner to that of \S~\ref{sec:photiso}, we wish to isolate the SN
component from our spectroscopic observations of GRB\,130702A. 
First, for absolute-flux calibration, we normalize our spectra to 
(uncorrected) broadband photometry at the appropriate epoch.  This accounts
for slit losses caused by variable seeing.

We next deredden our spectra of SN 2013dx in an analogous manner to 
that of \S~\ref{sec:photiso}.  This includes contributions from both the 
Milky Way [$E(B-V) = 0.038$\,mag] and the host galaxy 
($A_{V,\mathrm{host}} = 0.1$\,mag).

To remove the afterglow contribution, we assume that the spectrum can
be described at all times (and frequencies) as a power law of the form
$f_{\nu}(t, \alpha) \propto t^{-\alpha} \nu^{-\beta}$, with $\alpha = 
1.25$ and $\beta = 0.52$ (\S~\ref{sec:photiso}).  We normalize this
function to our (extinction-corrected) broadband photometry, and subtract
the appropriate power-law model at the epoch of each spectrum.

Finally, we fit the LMI/DCT late-time photometry to a variety of 
template galaxies from \citet{kcb+96}.  Similar to D15, we find that the
best-fit template is a starburst galaxy, and we adopted this 
(appropriately normalized) as the host contribution to the spectra.
We take this approach instead of using our final Keck/LRIS spectrum 
because of the blue-shutter failure.

After these corrections, only the SN component remains.  
Figure~\ref{fig:spectra} displays the resulting spectra of SN 2013dx after 
smoothing.  We exclude the first spectrum ($\Delta t = 1.17$\,d) because 
it is completely afterglow dominated.  In addition, we manually excise
nebular emission lines from the host galaxy of SN 2013dx.

\subsection{Comparison with Other Type Ic-BL SNe}
\label{sec:IcBL}

The early-time spectra of SN 2013dx are fairly featureless, but after a week,
broad ($v \approx 3 \times 10^{4}$\,km\,s$^{-1}$) features appear.  Together with 
the lack of obvious H and He emission, this leads us to classify SN 2013dx
as a broad-lined Type Ic SN (Ic-BL), as has been the case for essentially
all well-studied GRB-associated SNe thus far (e.g., \citealt{Woosley:2006}).

In Figure~\ref{fig:spectramax}, we plot the spectrum of SN 2013dx obtained
around maximum light ($\Delta t = 14.2$\,d) and at late times ($\Delta t = 33.3$\,d) 
with pseudocontinuum removed, bandpass filtered, scaled, and binned to 
a common logarithmic wavelength scale along with mean spectra of Type Ic-BL SNe both with 
and without GRBs from \citet{Modjaz:2015}.  We note that the mean spectra 
include spectra of SN 2013dx from D15, but this is one of many objects.  
The absorption features from mean Type Ic-BL SNe both with and without GRBs 
align well with SN 2013dx absorption features.  This indicates that SN 2013dx has similar 
photospheric velocities as other Type Ic-BL SNe. 

At maximum light SN 2013dx has a similar blueshifted, broad 
\ion{Si}{2} $\lambda$6355 line as both SNe~Ic-BL with and without GRBs. 
The blended \ion{Fe}{2} absorption feature around a blueshifted wavelength 
of 4800\,\AA\ is similar to that of SNe~Ic-BL with GRBs but is weaker than
that of SNe~Ic-BL without 
GRBs.  SN 2013dx has a stronger \ion{Ca}{2} absorption 
feature around 7900\,\AA\ and weaker \ion{O}{1} absorption feature 
around 7200\,\AA\ than most SNe~Ic-BL both with and 
without GRBs. 

At later epochs, SN 2013dx does not deviate from the mean SN~Ic-BL 
both with and without GRBs except beyond 8200\,\AA.  However, 
the relative variation from the continuum seems weaker than the mean 
spectra of both SNe~Ic-BL with and without GRBs.  SN 2013dx has a similar 
\ion{Ca}{2} absorption feature around 7900\,\AA\ as SNe~Ic-BL with 
GRBs but weaker than that of SNe~Ic-BL without GRBs.

To search for other similar objects in the literature, we use the 
cross-correlation tool SN Identification code (SNID; \citealt{Blondin:2007}).
Several SNe~Ic-BL that were not associated with GRBs, such as 
SN 1997ef \citep{inn+00} and SN 2007I \citep{CBET.808, Modjaz:2014}, also provide
good matches to SN 2013dx.  In addition, D15 highlight similarities to 
the energetic SN 2010ah (PTF10bzf; \citealt{cof+11,mwp+13}).

\subsection{Photospheric Velocity Measurements}
\label{sec:vel}

In order to estimate the photospheric velocity of SN 2013dx, 
we measure the velocity of the most prominent spectral feature,
the  \ion{Si}{2} 6355\,\AA\ absorption line.  We employ a 
fitting code in {\tt IDL} that removes a linear pseudocontinuum and fits a 
Gaussian to the absorption line (see \citealt{Silverman:2012} and 
\citealt{Silverman:2015} for a detailed description of the code).  
Table \ref{tab:pvel} displays the inferred velocities for each spectrum.  
Our results are also consistent with those reported by D15.
The first three spectra ($\Delta t = 1.17$, 3.25, and 6.22\,d) are too 
noisy for reliable velocity measurements.  

We note that \citet{Parrent:2015} suggests the absorption 
feature at $\sim6100$\,\AA, normally identified as \ion{Si}{2} 6355\,\AA,
may be instead associated with H$\alpha$.  Therefore, we compare 
mean SN~Ic-BL photospheric velocities measured using \ion{Fe}{2} 5169\,\AA\
\citep{Modjaz:2015} at maximum light.  We find that our measurements are 
consistent with the measurements from the less-contaminated 
\ion{Fe}{2} 5169\,\AA.

\begin{deluxetable}{cc}
\tabletypesize{\footnotesize}
\tablecolumns{2}
\tablewidth{150pt}
\tablecaption{Velocity of Si~II $\lambda$6355}
\tablehead{
\colhead{$\Delta t$} & \colhead{Velocity}  \\
\colhead{(d)} & \colhead{(km\,s$^{-1}$)} }
\startdata
9.3	&  28,100 $\pm$ 1000\\
11.3	&  25,200 $\pm$ 500\\
14.2	&  21,300 $\pm$ 500\\
31.3	&  11,700 $\pm$ 500\\
33.3	&  10,800 $\pm$ 500
\enddata
\tablecomments{In observer frame. The reported uncertainties come from fitting the Si~II $\lambda$6355 
absorption feature with a single Gaussian function and do not include errors from potential blending. }
\label{tab:pvel}
\end{deluxetable}

\subsection{Line Identification}
\label{sec:lines}

We use SYN++ \citep{Thomas:2011} to help identify the ions present in our spectra of SN 2013dx. SYN++ is derived from SYNOW \citep{Fisher:1997}, which uses the Sobolev approximation \citep{Sobolev:1960, Castor:1970, Jeffery:1989} to produce synthetic spectra of SNe during the photospheric phase. SYN++ assumes that spectral lines are formed via resonance scattering above a sharp photosphere. The location of the photosphere is expressed in velocity coordinates as $v_{\mathrm{ph}}$ (in km\,s$^{-1}$) and takes into account the homologous expansion of the ejecta.

The optical depths for each species must also be input and line strengths are computed assuming Boltzmann excitation (i.e., local thermodynamic equilibrium, LTE) using a specified excitation temperature $T_{\mathrm{exc}}$ (in K). Non-LTE effects are partially taken in account by allowing different $T_{\mathrm{exc}}$ values for each species, all of which can be different from the photospheric temperature $T_{\mathrm{phot}}$. The latter is used only in computing the blackbody radiation emitted by the photosphere.  
All our SYN++ fits are computed with all ions turned on simultaneously with a blackbody.

\begin{figure}[ht]
\centering
\begin{minipage}[b]{1.0\linewidth}
	\includegraphics[width = 1\linewidth]{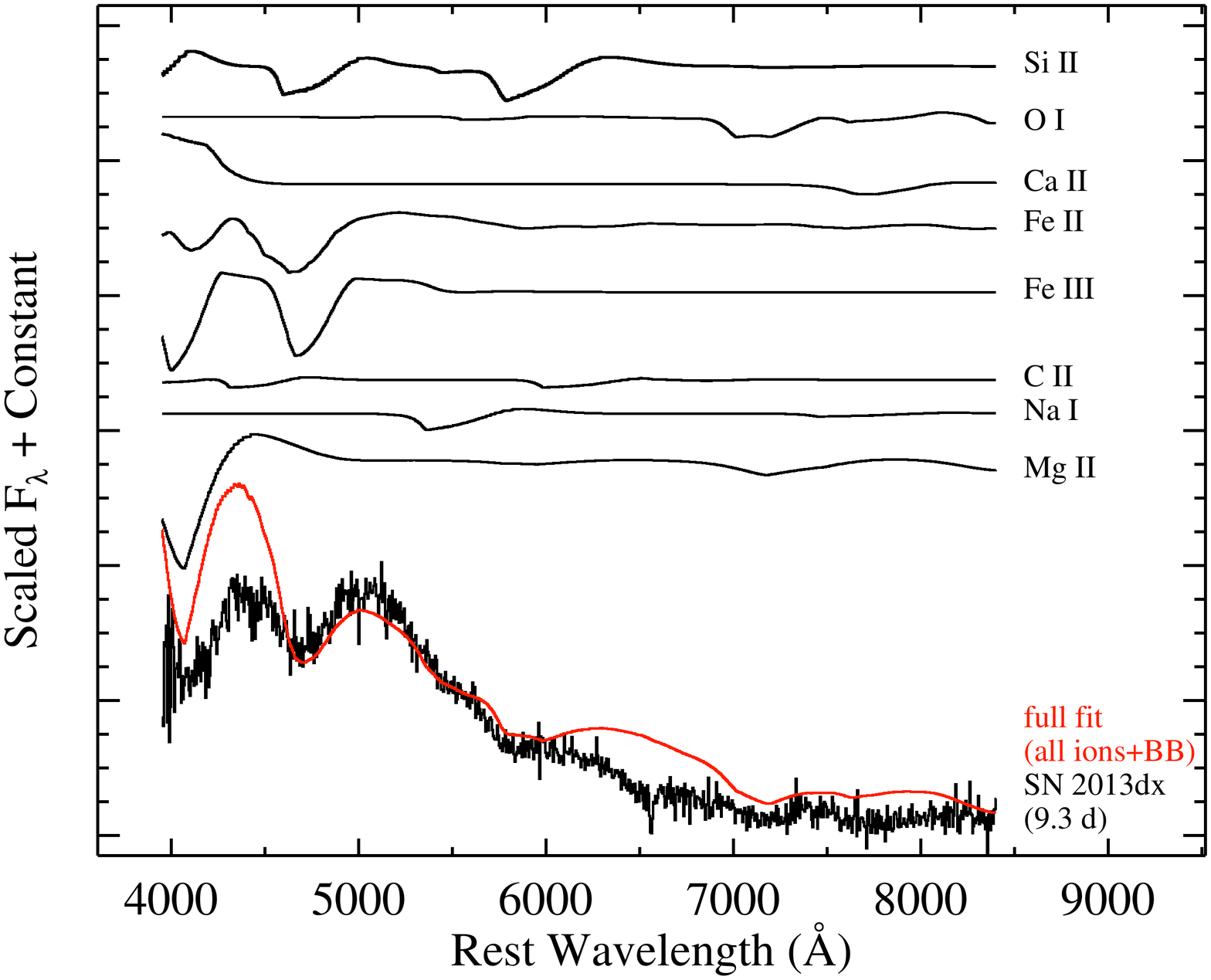}
\end{minipage}

\begin{minipage}[b]{1.0\linewidth}
	\includegraphics[width = 1\linewidth]{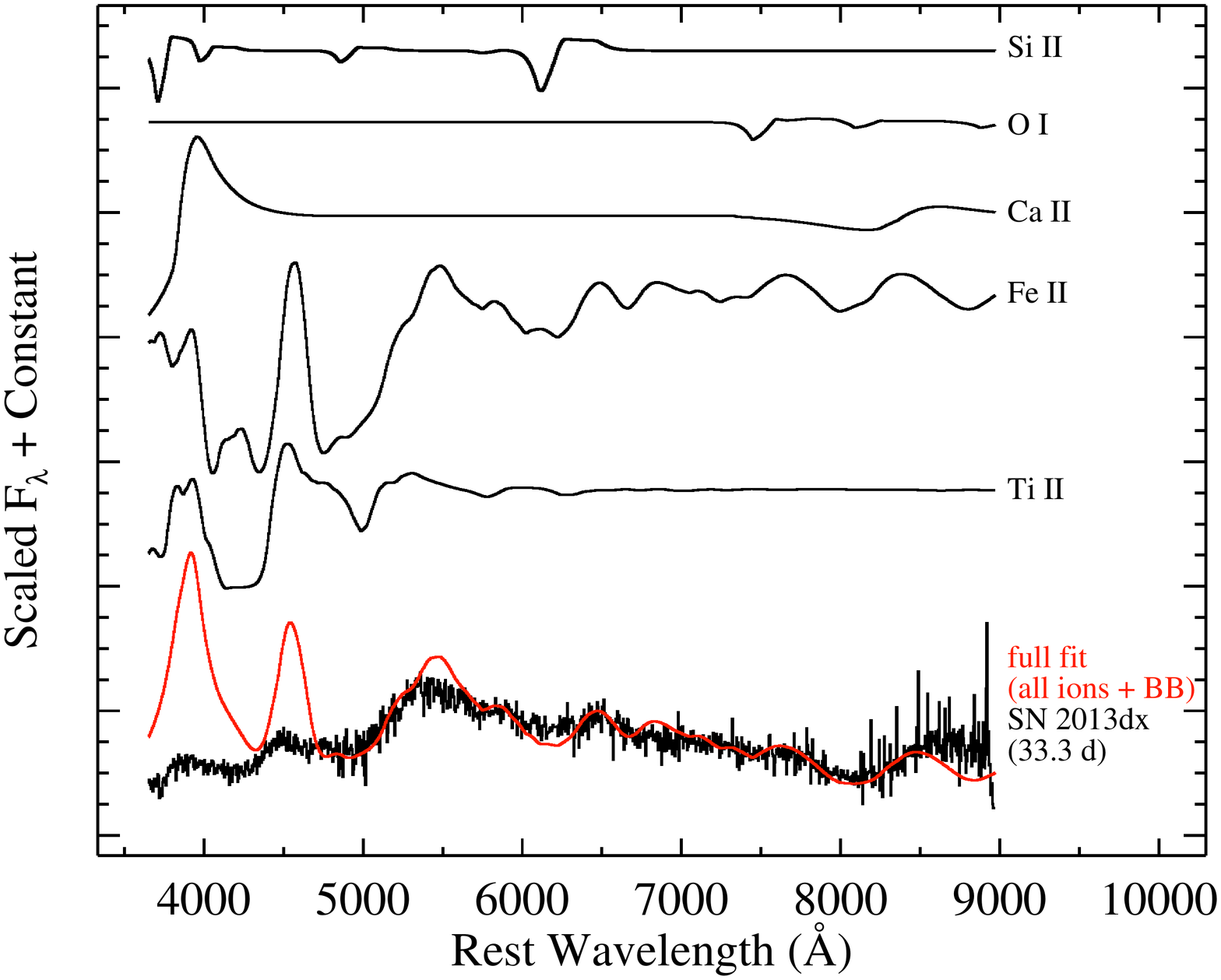}
\end{minipage}

\caption{ SYN++ fits to the 9.3 d (top) and 33.3 d (bottom) spectra of SN 2013dx. The spectrum of each individual ion is labeled. Their sum -- plus a 16,000\,K (top) and 9000\,K (bottom) blackbody -- is plotted in red on top of the actual observed spectra of SN 2013dx (binned to 6\,\AA\ per pixel).}
\label{fig:syn}
\end{figure}

We attempt to model the major spectral features of SN 2013dx at $\Delta t = 9.3$ and 33.3\,d and look for evolution during the photospheric phase (see Figure \ref{fig:syn}). At 9.3\,d after the burst, SYN++ indicates a photospheric velocity of 30,000\,km\,s$^{-1}$, an outer velocity of the line-forming region of about 90,000\,km\,s$^{-1}$, and an estimated photospheric temperature of 16,000\,K. The spectrum contains absorption from O~I, Si~II, and Fe~II, and possible absorption signatures of Fe~III, Mg~II, C~II, Ca~II, and Na~I.  We caution that because of
the relatively uncertain line identifications, the derived velocities
are robust indicators only for the lines of Fe~II and Si~II.

By 33.3\,d after the burst, according to our second SYN++ fit, the SN ejecta have slowed down and cooled off significantly. This model indicates approximate values for the photospheric velocity, outer velocity, and photospheric temperature of 11,000\,km\,s$^{-1}$, 60,000\,km\,s$^{-1}$, and 9000\,K, respectively. The majority of the absorption in this spectrum is likely produced by Fe~II and Ti~II, though there is
some evidence of Si~II, Ca~II, and possibly O~I as well. While the fit to this spectrum at wavelengths below $\sim4700$\,\AA\ is not perfect, the broad peaks and troughs roughly match.  This part of the spectrum is notoriously difficult to model owing to hundreds of overlapping spectral features, mostly from iron-group elements.

%%%%%%%%%%%%%%%%%%%%%%%%%%%%%%%%%%%%%%%%%%%%%%

\section{Supernova Explosion Parameters}
\label{sec:SNparams}

\subsection{Derived Parameters}
\label{sec:deriveSNparams}

We model the basic explosion parameters of SN 2013dx by fitting its bolometric light curve with the Type I SN analytical model of \citet{Arnett:1982} and \citet{Valenti:2008}.  This model assumes (1) homologous expansion of the ejecta, (2) spherical symmetry, (3) all $^{56}$Ni is located at the center of explosion and no mixing, (4) radiation-pressure dominated ejecta, (5) the initial radius before explosion is small, (6) the diffusion approximation is appropriate for photons (i.e., the ejecta are in the photospheric phase), and (7) a single opacity over the duration of the explosion.  

The peak luminosity correlates with the mass of $^{56}$Ni ($M_{\mathrm{Ni}}$), while the light-curve shape is determined by the total ejecta mass ($M_{\mathrm{ej}}$) and the ejecta kinetic energy ($E_{\mathrm{K}}$).  We can break the degeneracy between $M_{\mathrm{ej}}$ and $E_{\mathrm{K}}$ with photospheric-velocity measurements from our optical spectra. 

The timescale of the light curve is given by
\begin{equation}\label{eq:valenti2}
\tau_m =  \bigg( \frac{\kappa}{\beta c} \bigg)^{1/2} \bigg( \frac{6M^3_{\mathrm{ej}}}{5E_{\mathrm{K}}} \bigg)^{1/4},
\end{equation}
where $\beta \approx$ 13.8 is an integration constant.  For a uniform density \citep{Arnett:1982}\footnote{\label{fn:arnett}Note there is a typo incorrectly stating $E_{\mathrm{K}} \approx \frac{5}{3} \frac{M_{\mathrm{ej}}v_{\mathrm{ph}} ^2}{2}$ in the original text that was corrected by \citet{Arnett:1996}.  This has been taken into account in Equations \ref{eq:valenti2} and \ref{eq:valenti3}.  This typo has been propagated throughout the literature.},
\begin{equation}\label{eq:valenti3}
E_{\mathrm{K}} \approx \frac{3}{5} \frac{M_{\mathrm{ej}}v_{\mathrm{ph}}^2}{2}.
\end{equation}
We assume $\kappa = 0.07$\,cm$^2$\,g$^{-1}$ to directly compare with the literature for other GRB-SNe (e.g., \citealt{Cano:2011}). 

We fit our quasi-bolometric light curve (\S~\ref{sec:bolometric}) with the Arnett-Valenti relation,
\begin{equation}\label{eq:valenti1}
\begin{split}
L_{\mathrm{ph}}(t)= & M_{\mathrm{Ni}}e^{-x^{2}} \\
		& \times \left [ (\epsilon_{\mathrm{Ni}}-\epsilon_{\mathrm{Co}})\int_{0}^{x}A(z)dz + \epsilon_{\mathrm{Co}}\int_{0}^{x}B(z)dz \right ],
\end{split}
\end{equation}
with
\begin{equation}
\begin{split}
A(z) &= 2ze^{-2zy+z^2}, \\
B(z) &= 2ze^{-2zy+2zs+z^2}, \\
x &\equiv t/ \tau_{\mathrm{m}}, \\
y &\equiv \tau_{\mathrm{m}}/(2\tau_{\mathrm{Ni}}),~{\rm and} \\
s &\equiv \tau_m(\tau_{\mathrm{Co}}-\tau_{\mathrm{Ni}})/(2\tau_{\mathrm{Co}}\tau_{\mathrm{Ni}}).
\end{split}
\end{equation}
The decay times of $^{56}$Ni and $^{56}$Co are $\tau_{\mathrm{Ni}} = 8.77$\,d and $\tau_{\mathrm{Co}} = 111.3$\,d, and the energies produced in one second by one gram of $^{56}$Ni and $^{56}$Co were taken as $\epsilon_{\mathrm{Ni}} = 3.90 \times 10^{10}$\,erg\,s$^{-1}$\,g$^{-1}$ and $\epsilon_{\mathrm{Co}} = 6.78 \times 10^{9}$\,erg\,s$^{-1}$\,g$^{-1}$ \citep{Sutherland:1984,Cappellaro:1997}.

From our spectra and light curves, the SN component was dominant starting at $\Delta t \approx 4$\,d (compare with SN 2010bh, where shock breakout was prominent out to 7\,d; \citealt{Cano:2011}).  The Arnett-Valenti relation assumes that the material is in the photospheric phase, which is no longer valid at $\Delta t \gtrsim 30$\,d.  Therefore, our fit only includes $4 \leq \Delta t \leq 30$\,d.  We find $M_{\mathrm{Ni}} = 0.37 \pm 0.01$\,M$_{\odot}$ and $\tau_{\mathrm{m}} = 11.35 \pm 0.17$\,d (statistical errors only).  Using $v_{\mathrm{ph}} = 21,300\, \text{km\,s}^{-1}$ from our spectral fit near peak (\S~\ref{sec:vel}), we calculate $M_{\mathrm{ej}} = 3.1 \pm 0.1$\,M$_{\odot}$ and $E_{\mathrm{K}} = (8.2 \pm 0.43) \times 10^{51}$\,erg.

\begin{deluxetable*}{llrccccc}
\tabletypesize{\footnotesize}
\tablecolumns{8}
\tablewidth{0pt}
\tablecaption{Physical Parameters of GRB-SNe}

\tablehead{
\colhead{GRB-SN} & \colhead{$z$} & \colhead{$E_{\gamma,{\rm iso}}$ }& \colhead{$v_{\mathrm{ph}}$} & \colhead{$M_{\mathrm{Ni}}$  }& \colhead{$M_{\mathrm{ej}}$  } & \colhead{$E_{\mathrm{K}}$}& \colhead{Reference}  \\
\colhead{} & \colhead{} & \colhead{(erg)} & \colhead{(km\,s$^{-1}$)} & \colhead{(M$_{\odot}$)} & \colhead{(M$_{\odot}$)} & \colhead{($10^{52}$ erg)} & \colhead{}    }

\startdata

GRB 980425/SN 1998bw	& 0.0085 & $(9.29 \pm 0.35) \times 10^{47}$ &  18,000 & $0.42 \pm 0.02$ & $6.80 \pm 0.57$ & $1.31 \pm 0.10$\textsuperscript{a}& (1), (2) \\
GRB 030329/SN 2003dh	& 0.1685 & $1.33 \times 10^{52}$ &  20,000 & $0.54 \pm 0.13$ & $5.06 \pm 1.65$ & $1.21 \pm 0.39$\textsuperscript{a} & (1), (2) \\
GRB 031203/SN 2003lw	& 0.105& $1.67^{+0.04}_{-0.10} \times 10^{50}$ &  18,000 & $0.57 \pm 0.04$ & $8.22 \pm 0.76$ & $ 1.59 \pm 0.15$\textsuperscript{a} & (1), (2)\\
GRB 060218/SN 2006aj	& 0.0335 & $4.33^{+0.41}_{-1.74} \times 10^{49}$ &  20,000 & $0.21 \pm 0.03$ & $2.58 \pm 0.55$ & $0.61 \pm 0.14$\textsuperscript{a} & (1), (2)\\
GRB 091127/SN 2009nz	& 0.49 & $(4.3 \pm 0.3) \times 10^{52}$ &  17,000 & $0.33 \pm 0.01$ & $4.69 \pm 0.13$ & $0.81 \pm 0.02$\textsuperscript{a} & (1), (3)\\
GRB 100316D/SN 2010bh$^b$ & 0.059 & $\ge (3.9 \pm 0.3) \times 10^{49}$	&  25,000 & $0.12 \pm 0.02$ & $2.47 \pm 0.23$ & $0.92 \pm 0.08$\textsuperscript{a} & (1), (4)\\
GRB 120422A/SN 2012bz$^b$  & 0.283 & $4.5 \times 10^{49}$	&  20,500 & $0.57 \pm 0.07$ & $6.10 \pm 0.49$ & $1.53 \pm 0.13$\textsuperscript{a} & (1), (5)\\
GRB 130427A/SN 2013cq & 0.3399 & $(9.6 \pm 0.04) \times 10^{53}$ &  32,000 & $0.28 \pm 0.02$ & $6.27 \pm 0.69$ & $6.39 \pm 0.70$ & (6)\\
GRB 130702A/SN 2013dx & 0.145 & $6.4^{+1.3}_{-1.0} \times 10^{50}$ &  21,300 & $0.37 \pm 0.01$ & $3.1 \pm 0.1$ & $0.82 \pm 0.04$ & (7)\\
GRB 140606B/iPTF14bfu	& 0.384 & $(3.47 \pm 0.02) \times 10^{51}$ &  19,820 & $0.42 \pm 0.17$ & $4.8 \pm 1.9$ & $1.1 \pm 0.7$\textsuperscript{a} & (8)

\enddata
\tablecomments{\textsuperscript{a}$E_{\rm K}$ originally calculated as $E_{\rm K} = \frac{M_{\rm ej}v_{\rm ph}^2}{2}$, scaled by a factor of 3/5 to directly compare with our values. All $E_{\gamma,{\rm iso}}$ values calculated over 1\,keV -- 10\,MeV, except those indicated by superscript ``b'' which are calculated over 15--150\,keV or superscript ``c'' which are calculated over 20--1400\,keV.  (1) \citet{Cano:2013}, (2) \citet{Kaneko:2007}, (3) \citet{Troja:2012}, (4) \citet{Starling:2011},  (5) \citet{Zhang:2012}, (6) \citet{xdl+13}, (7) this work, (8) \citet{cdp+15}.}
\label{tab:energetics}
\end{deluxetable*}

\subsection{Comparison with Other GRB-SNe}
\label{sec:compareGRBSNe}

\begin{figure*}[ht]
\centering
\begin{minipage}[b]{0.48\linewidth}
	\includegraphics[width=1\linewidth]{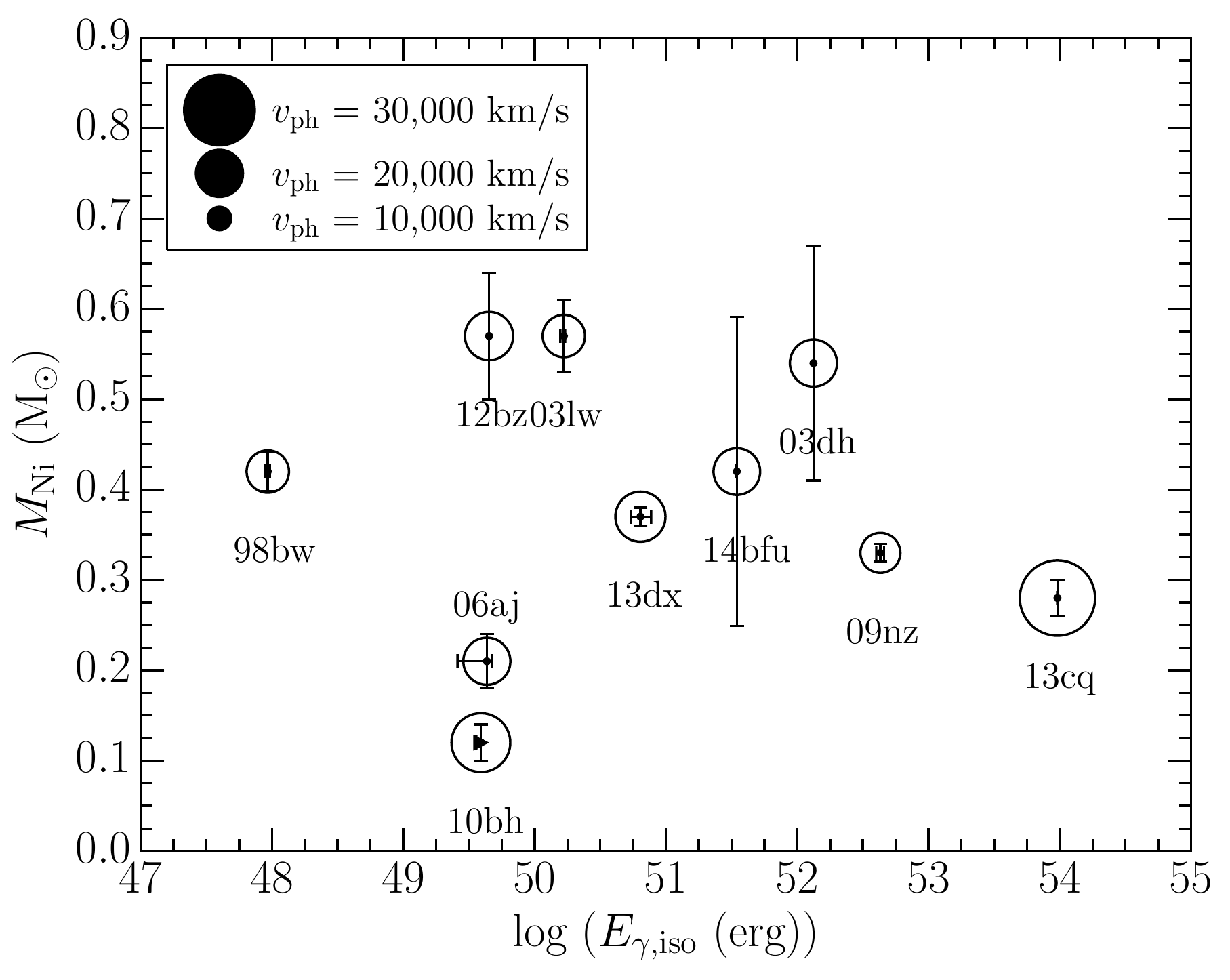}
\end{minipage}
\quad
\begin{minipage}[b]{0.48\linewidth}
	\includegraphics[width = 1\linewidth]{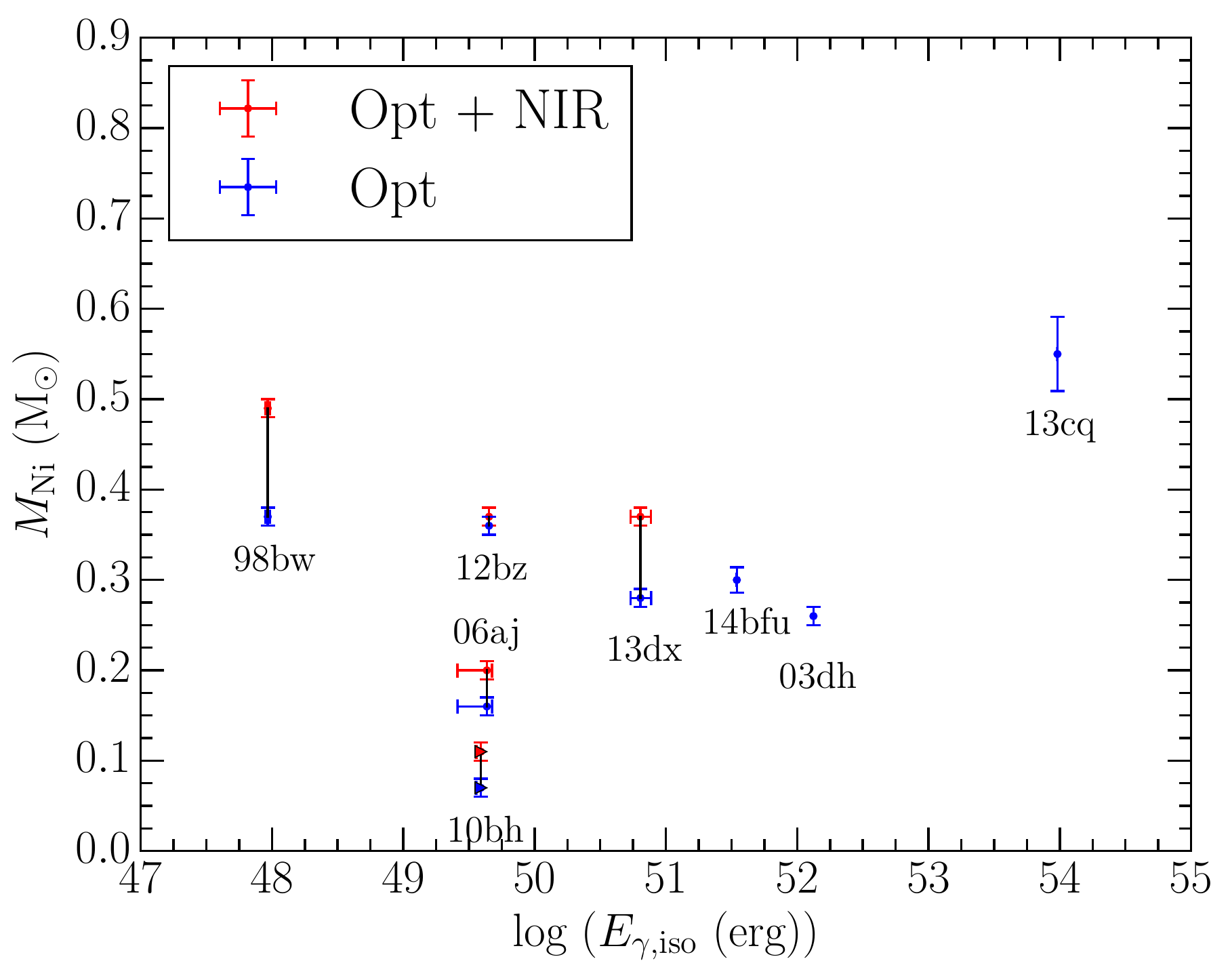}
\end{minipage} 

\caption{ \textit{(left)} Comparison of SN explosion parameters from template GRB-SNe and SN 2013dx from Table \ref{tab:energetics}.  The size of the points indicates $v_{\mathrm{ph}}$ near the SN brightness peak.  There does not appear to be any correlation between $M_{\mathrm{Ni}}$ and $E_{\gamma,{\rm iso}}$ or $M_{\mathrm{Ni}}$ and $v_{\mathrm{ph}}$. 
\textit{(right)} Comparison of SN explosion parameters using photometric data reported in the literature with our quasi-bolometric fitting procedure.  This avoids using SN 1998bw as a template for other GRB-SN bolometric fits (see text for details).  Blue points are fit to only optical data, red points are fit to optical and NIR data. Black line connects points that have both.}
\label{fig:SNpar} 
\end{figure*}

\begin{figure*}[ht]
\centering
	\includegraphics[width = 0.5\linewidth]{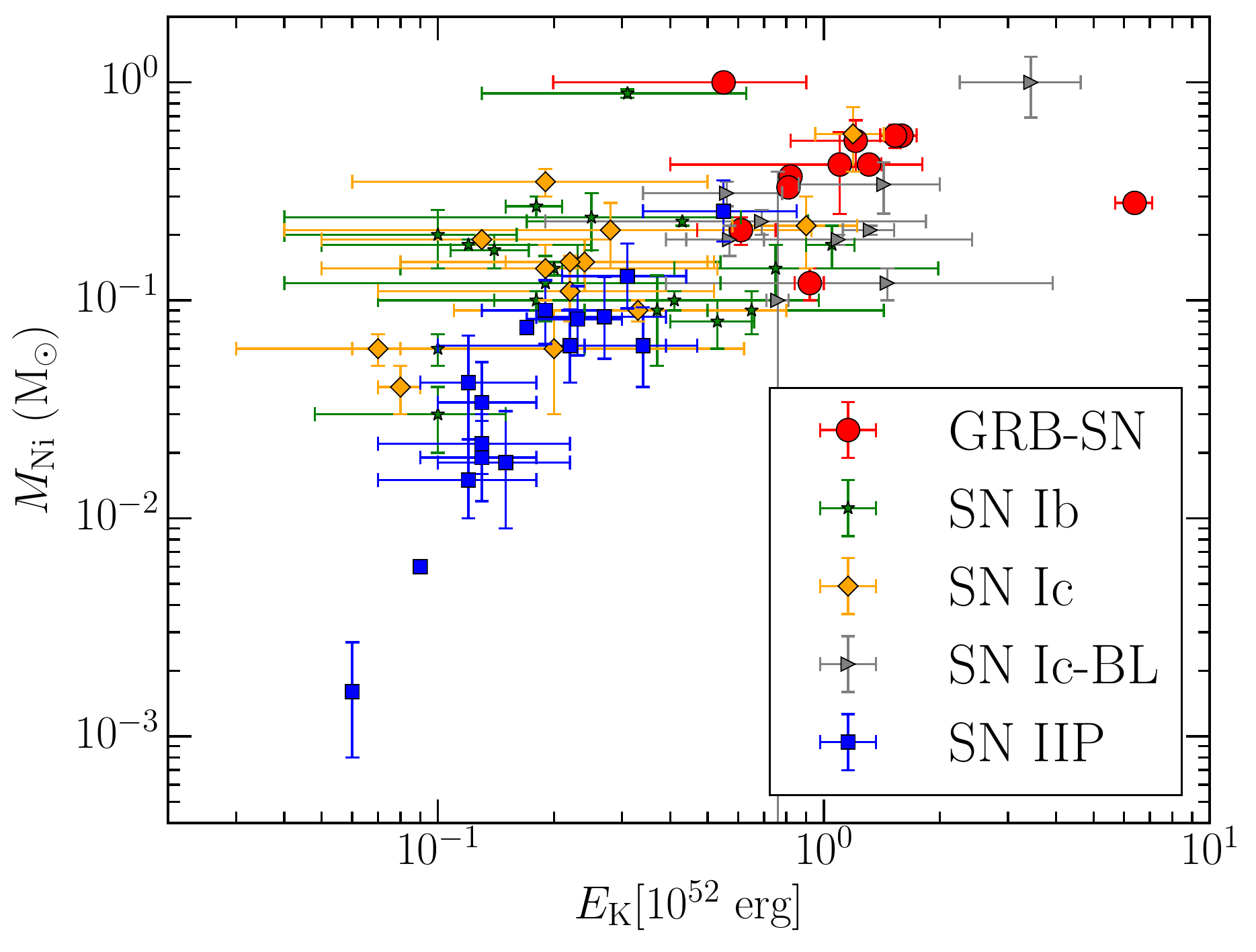}
\caption{There is a clear correlation between explosion energy, $E_{\rm K}$, and $M_{\mathrm{Ni}}$ for SNe~Ib, Ic, Ic-BL (\citealt{Cano:2013}, and references therein), IIP (\citealt{Hamuy:2003}, and references therein), and GRB-SNe from Table \ref{tab:energetics}. See \citet{f97} for a review of SN classification. }
\label{fig:SNparEk} 
\end{figure*}

We compare our bolometric light curve of SN 2013dx with that of other spectroscopically confirmed GRB-SNe in Figure \ref{fig:bolometric} (SN 1998bw, \citealt{Galama:1998}; SN 2003lw, \citealt{Mazzali:2006}; SN 2003dh, \citealt{Deng:2005};  SN 2006aj, \citealt{pmm+06}; SN 2009nz, \citealt{Olivares:2015}, SN 2010bh, \citealt{ogs+12}; SN 2012bz: \citealt{Melandri:2012} and \citealt{Schulze:2014}; SN 2013cq: \citealt{mpd+14}; iPTF14bfu: \citealt{cdp+15}).  We note that the NIR contribution to the bolometric luminosity for SN 2012bz is assumed to be the same as that observed for SN 2010bh.
Although these GRB-SN bolometric light curves cover different wavelength ranges, we can get a sense of the light-curve evolution.  SN 2013dx most closely matches the light-curve shape of SN 2012bz, but with the caveat that SN 2013dx has a steeper rise than SN 2012bz.  

After 30 days, SN 2013dx appears to drop in luminosity rapidly.  This is unlike the three bursts with late-time coverage -- SN 2003lw, SN 2003dh, and SN 1998bw -- seen particularly well juxtaposed against SN 1998bw, which has extensive observations out to hundreds of days.  This drop in luminosity is not from underestimating the NIR contribution at late times; we still observe a rapid drop in luminosity at 30\,d if we continue the monotonically increasing NIR ratio function from \S~\ref{sec:bolometric} instead of adding a flat NIR contribution of 23\% after 29\,d. 

In order to compare the derived properties of SN 2013dx with a broader sample of
GRB-associated SNe, we use the derived values for $M_{\mathrm{Ni}}$, $M_{\mathrm{ej}}$,
and $E_{\mathrm{K}}$ for all well-sampled events from \citet{Cano:2013}.  The authors
fit a template of SN 1998bw to determine the appropriate stretch and scale parameters
(e.g., \S~\ref{sec:sn98bw}).  Using average $s$ and $k$ values for each burst, 
the authors then fit a scaled version of the $UBVRIJH$ SN 1998bw light curve 
to the Arnett-Valenti model to derive the explosion parameters.  This method has
the benefit of (effectively) uniform wavelength coverage, even for events that were
only observed in a few filters.  However, the primary drawback is the assumption
that all bolometric light curves are well fit by an appropriately scaled version
of SN 1998bw.  As evidenced by Figure~\ref{fig:sk}, this assumption breaks down 
at the very least for the redder filters for SN 2013dx (see also \citealt{lbj+14}).
Nonetheless, the explosion parameters for all well-studied GRB-associated SNe 
derived in this manner\footnote{We note that \citet{Cano:2013} and \citet{cdp+15} 
assume that $E_{\mathrm{K}} = \frac{M_{\rm ej}v_{\rm ph}^2}{2}$, so we scaled their 
reported $E_{\mathrm{K}}$ values by a factor of 3/5.} are presented in 
Table~\ref{tab:energetics}.

To avoid any biases introduced by GRB-SNe that are not 
well matched with SN 1998bw, we also create our own quasi-bolometric light curves 
by compiling photometry from the literature  (Table \ref{tab:photoref}).  We fit these light curves
with the Arnett-Valenti relation described in \S~\ref{sec:deriveSNparams} and break our 
results into those events with only optical datasets and those with optical and NIR photometry in Figure~\ref{fig:SNpar} (right panel).  This provides an estimate
of the fundamental explosion parameters that is independent of any assumed
similarity with SN 1998bw.    

We plot the inferred $M_{\mathrm{Ni}}$ as a function
of both the isotropic prompt gamma-ray energy release (Figure~\ref{fig:SNpar}) and 
 the derived SN kinetic energy ($E_{\mathrm{K}}$; in Figure~\ref{fig:SNparEk}).

\begin{deluxetable*}{lcl}
\tabletypesize{\footnotesize}
\tablecolumns{3}
\tablewidth{0pt}
\tablecaption{Photometry References for SN 1998bw-Independent Bolometric Light Curves}
\tablehead{
\colhead{GRB-SN} & \colhead{Filter Coverage} & \colhead{References}
}
\startdata
GRB 980425/SN 1998bw	& $UBVRIJHK$ & \citet{Clocchiatti:2011} \\
GRB 030329/SN 2003dh	& $UBVR$ & \citet{Deng:2005}\\
GRB 031203/SN 2003lw\textsuperscript{a}	& --- & \\
GRB 060218/SN 2006aj	& $BVRIJHK$ & \citet{fkz+06, kmb+07}\\
GRB 091127/SN 2009nz\textsuperscript{a}	& --- & \\
GRB 100316D/SN 2010bh\textsuperscript{b}   & $grizJH$ & \citet{ogs+12} \\
GRB 120422A/SN 2012bz  & $grizJ$ & \citet{Schulze:2014}\\
GRB 130427A/SN 2013cq\textsuperscript{c}  & $BVRI$ &\citet{mpd+14} \\
GRB 130702A/SN 2013dx  & $grizyJ$ & This work\\
GRB 140606B/iPTF14bfu	& $griz$ & \citet{cdp+15}
\enddata
\tablecomments{\textsuperscript{a}Simultaneous epochs near peak were limited and were insufficient for our analysis tools. 
\textsuperscript{b}Used lower host-galaxy extinction from \citet{Cano:2011} and \citet{bps+12}.
\textsuperscript{c}Poor fit using our analysis tools; we fit to the bolometric curve provided by E. Pian (private communications). }
\label{tab:photoref}
\end{deluxetable*}

\subsection{Caveats and Conclusions}
\label{sec:caveats}

From Figure~\ref{fig:SNpar}, it is clear
that there is no correlation between $M_{\mathrm{Ni}}$ and the prompt energy
release using either method for calculating SN explosion parameters of 
other GRB-SNe.  For example, SN 2013dx has a comparable mass of synthesized $^{56}$Ni as the
subluminous GRB\,980425 / SN 1998bw and the extremely luminous GRB\,130427A / 
SN 2013cq \citep{ltf+14,mpd+14}.  Even if we were to apply a beaming correction,
GRB\,130427A would still have $E_{\gamma}$ several orders of magnitude larger
than GRB\,980425 \citep{pcc+14}, but comparable $M_{\mathrm{Ni}}$.  Similarly,
there is no clear correlation between $M_{\mathrm{Ni}}$ and photospheric
velocity at peak.  Numerical simulations unambiguously predict that the mass
of synthesized $^{56}$Ni should be correlated with the degree of asymmetry 
in the explosion \citep{gdr+14,Umeda:2008}; to the extent that our models 
faithfully reproduce the relevant (global) SN explosion parameters, this 
result is clearly not borne out by the data.

On the other hand, the explosion energy \textit{of the SN ejecta} is clearly
correlated with $M_{\mathrm{Ni}}$, particularly when including other core-collapse
events.  As shown by previous authors (e.g., \citealt{Cano:2013} and \citealt{lbj+14}
for recent compilations), GRB-associated SNe on average have a higher mass of 
synthesized $^{56}$Ni and larger kinetic energies than any other class of
core-collapse SNe (except perhaps the superluminous SNe; 
\citealt{g12}).  That said, the SN explosion energies are typically narrowly
clustered and do not appear to significantly exceed $10^{52}$\,erg, consistent with
(perhaps even indicative of) a magnetar origin for these events \citep{mmw+14}.
SN 2013cq (associated with GRB\,130427A) appears to be a significant outlier
in terms of its inferred $E_{\mathrm{K}}$, however, which remains to be fully
explained.

D15 created a bolometric light curve over the range 3000--10,000\,\AA\ extrapolated from $u^{\prime}$ and $i^{\prime}$.  We examined the bolometric light curve from D15 and found that our peak luminosity is consistent with theirs when accounting for our additional NIR coverage.   D15 report $M_{\mathrm{Ni}} \approx 0.2\, {\rm M}_{\odot}$, a factor of two lower 
than our quoted value.  On the other hand, they derive a total ejecta mass
($M_{\mathrm{ej}} \approx 7 \pm 2$\,M$_{\odot}$) and a SN kinetic energy 
($E_{\mathrm{K}} \approx (3.5 \pm 1.0) \times 10^{52}$ erg) approximately a factor 
of two and four (respectively) larger than those presented here.   D15 scale numerical simulations of the similarly shaped SN 2003dh \citep{Mazzali:2006} to estimate SN explosion parameters, as opposed to using an analytical model 
(e.g., Arnett-Valenti) as adopted in this paper.  

Since $M_{\mathrm{Ni}}$ is closely related to the peak luminosity, we believe that our reported $M_{\mathrm{Ni}}$ estimate is more accurate than the value D15 report. When NIR contributions are included, SN 2013dx has a similar peak luminosity as SN 1998bw.  We therefore expect SN 2013dx to have a similar $^{56}$Ni mass as SN 1998bw. Both numerical simulations and analytical models produce $M_{\mathrm{Ni}} \approx 0.4\,{\rm M}_{\odot}$ for SN 1998bw \citep{Mazzali:2006, Cano:2013}.    

The discrepancy in ejecta mass and kinetic energy is caused mainly by different opacity assumptions.   \citet{Mazzali:2006} assumed an opacity, $\kappa = 0.5\, Y_e$\,cm$^2$\,g$^{-1}$, where $Y_e$ is the number of electrons per baryon. We assume the authors used $Y_e = 0.46$ for iron and recalculate our Arnett-Valenti fit.  With an opacity of $\kappa = 0.02$\,cm$^2$\,g$^{-1}$, we report $M_{\mathrm{ej}} = 9.2 \pm 0.2$\,M$_{\odot}$ and $E_{\mathrm{K}} = (2.5 \pm 0.1) \times 10^{52}$ erg.  We also note that from our fit of SN 1998bw in \S \ref{sec:sn98bw}, we can see that the light-curve evolution of SN 1998bw does not match that of SN 2013dx well (see Figure \ref{fig:sk}); hence, the different values of $M_{\mathrm{ej}}$ and
$E_{\mathrm{K}}$ are not unexpected.  

We note that opacity greatly affects $M_{\mathrm{ej}}$ and $E_{\mathrm{K}}$, but does not affect $M_{\mathrm{Ni}}$.  \citet{Wheeler:2015} found that a conflict exists when comparing properties determined by fitting the peak (using Arnett-Valenti methods) to those determined from fitting the late-time tail (using methods from \citealt{Clocchiatti:1997}).  This conflict is partially resolved by using a mean opacity determined from both peak and late-time tail parameters.  \citet{Wheeler:2015} find that in general, the opacity is often overestimated in the literature and the typical mean opacity is $\kappa \approx 0.01 \text{ cm}^2 \text{ g}^{-1}$.  In view of this potential discrepancy, we report values for $\kappa = 0.01 \text{ cm}^2 \text{ g}^{-1}$ ($M_{\mathrm{ej}} \approx 21 \,{\rm M}_{\odot}$, $E_{\mathrm{K}} \approx 6 \times 10^{52}$\,erg) and $\kappa = 0.1 \text{ cm}^2 \text{ g}^{-1}$ ($M_{\mathrm{ej}} \approx 2 \,{\rm M}_{\odot}$, $E_{\mathrm{K}} \approx 6 \times 10^{51}$\,erg) to draw attention to the spread in $M_{\mathrm{ej}}$ and $E_{\mathrm{K}}$ from this variable.

We also caution that numerical simulations of jet-driven SNe (e.g., \citealt{gdr+14,mwp+13,Umeda:2008}) imply that
the distribution of $^{56}$Ni is likely to be highly asymmetric.  The derived ejecta
mass may therefore be biased by line-of-sight effects, and not representative of the
total mass ejected in the explosion.

%%%%%%%%%%%%%%%%%%%%%%%%%%%%%%%%%%%%%%%%%%%%%%
\section{Discussion and Summary}

We present extensive optical and NIR photometry of GRB\,130702A/SN 2013dx spanning 1--63\,d after the gamma-ray trigger, and optical spectra covering 1--33\,d after the trigger. At $z = 0.145$, GRB\,130702A/SN 2013dx is sufficiently close to clearly detect and model the underlying SN component that emerged a week after the burst.  

We isolate the SN component and present multi-band light curves, a quasi-bolometric ($g^{\prime}r^{\prime}i^{\prime}z^{\prime}yJ$) light curve, and spectra of SN 2013dx. Detection of the broad \ion{Si}{2} $\lambda 6355$ absorption line at  velocities approaching $3 \times 10^{4}$\,km\,s$^{-1}$, combined with the absence of H and He features, indicates that SN 2013dx is a broad-lined SN~Ic.  We estimate the SN explosion parameters using the Arnett-Valenti analytical relation and infer $M_{\mathrm{Ni}} = 0.37 \pm 0.01$\,M$_{\odot}$, $M_{\mathrm{ej}} = 3.1 \pm 
0.1$\,M$_{\odot}$, and $E_{\mathrm{K}} = (8.2 \pm 0.4) \times 10^{51}$\,erg.

Our analysis allows us to compare SN 2013dx with other GRB-SNe, as well as other
core-collapse SNe (those of identical spectral type and not).  This is of particular interest because GRB 130702A is of intermediate $E_{\gamma,\mathrm{iso}}$, between low-luminosity and cosmological GRBs.  There seems to be no clear relation between $M_{\mathrm{Ni}}$, $M_{\mathrm{ej}}$, or $E_{\mathrm{K}}$ with GRB isotropic energy (Figure~\ref{fig:SNpar}), even when considering beaming corrections.  
The SN appears to not be imprinted with any information about the formation of the relativistic jet aside from the high photospheric velocity and lack of H and He that allows us to classify all GRB-SNe as Type Ic-BL.  This is somewhat puzzling, given the
predictions of a correlation between the degree of asymmetry and mass of synthesized
$^{56}$Ni for jet-driven explosions.  We caution that the Arnett-Valenti relations we use 
to derive $M_{\mathrm{Ni}}$ assume spherical symmetry and this assumption may account for some but not all of the scatter in Figure~\ref{fig:SNpar}. On the other hand, our observations do provide support for predictions that $M_{\mathrm{Ni}}$ should be strongly correlated with
the kinetic energy of the SN itself.

Spectroscopically, SN 2013dx resembles both other GRB-SNe like SN 2006aj and SN 1998bw, as well as non-GRB SNe~Ic-BL such as SN 1997ef, SN 2007I, and SN 2010ah.  In terms of light curves, SN 2013dx most closely matches the evolution of SN 2012bz, associated with an intermediate GRB, but has a similar peak luminosity as SN 1998bw, associated with a low-luminosity GRB.  Direct comparison of the light-curve evolution between SN 2013dx and SN 1998bw indicates that SN 2013dx has a quicker rise time than SN 1998bw.  The faster rise time may suggest that SN 2013dx has a steeper distribution of $^{56}$Ni in the outer layers of the star (i.e., less mixing) than SN 1998bw (\citealt{Piro:2013}, \citealt{Dessart:2012}).  

Late-time observations several months after the burst can test asymmetry effectively both in photometry 
\citep{Wheeler:2015} and spectroscopy \citep{Mazzali:2005, Maeda:2002,Milisavljevic:2015}.  We do not have enough late-time 
observations of SN 2013dx to conduct these effective asymmetry tests, but we strongly encourage late-time follow-up data
for GRB-SNe when possible.

Finally, we suggest two potential avenues for future study, especially
with respect to GRB\,130702A / SN 2013dx.  Detailed numerical modeling of the SN ejecta (e.g., \citealt{Mazzali:2006} for SN 1998bw), specifically tailored to the light
curves and spectra of SN 2013dx (instead of simply scaling results from previous
simulations), should help to improve the accuracy of estimates of the fundamental
SN explosions parameters.  In addition, a broadband study of the afterglow 
emission, in particular incorporating radio wavelengths, would enable a 
much-improved estimate of the properties of the fastest-moving ejecta.  This would greatly
assist in placing GRB\,130702A in the context of other relativistic explosions, 
specifically how the explosion energy is partitioned with respect to ejecta 
velocity (e.g., \citealt{mms+14}).

%%%%%%%%%%%%%%%%%%%%%%%%%%%%%%%%%%%%%%%%%%%%%%

\section*{Acknowledgements}

We gratefully acknowledge S. Schulze, F. Olivares E., A. Melandri, M. Modjaz, Y. Liu, and E. Pian for generously sharing their raw data, which significantly improved the analysis for this paper.  We also thank J. Lyman, M. Modjaz, R. Thomas, and the anonymous referee for useful feedback on the manuscript.

This work was supported by the National Aeronautics and Space
Administration (NASA) Headquarters under the NASA Earth and Space Science Fellowship Program (Grant NNX12AL70H to V.T.). 
V.T. and S.V. were partially supported by NSF/ATI grant 1207785.  The research of A.V.F.'s group at UC Berkeley 
has been funded by NSF grant AST-1211916, Gary and Cynthia Bengier, the Richard and Rhoda Goldman Fund, the TABASGO Foundation, and the Christopher R. Redlich Fund.
J.M.S. is supported by an NSF Astronomy and Astrophysics Postdoctoral Fellowship under award AST-1302771.  
A.G.Y. is supported by the EU/FP7 via ERC grant no. 307260, the Quantum
Universe I-Core program by the Israeli Committee for planning and
budgeting and the ISF; by Minerva and ISF grants; by the Weizmann-UK
``making connections'' program; and by Kimmel and ARCHES awards.
A.C. acknowledges support from the NASA-\textit{Swift} GI program via grants 13-SWIFT13-0030 and 14-SWIFT14-0024.  E.T. acknowledges support for this project under the {\it Fermi} Guest Investigator Program.  The work of D.S. was carried out at the Jet Propulsion Laboratory, California Institute of Technology, under a contract with NASA.  We also acknowledge the help of K. Markey, E. Alduena, A. Alduena, and S. Kuo from Walden School for their help with the Palomar observations on 2013 July 8.

We thank the RATIR project team and the staff of the Observatorio Astron\'{o}mico Nacional on Sierra San Pedro M\'{a}rtir. RATIR is a collaboration between the University of California, the Universidad Nacional Auton\'{o}ma de M\'{e}xico, NASA Goddard Space Flight Center, and Arizona State University, benefiting from the loan of an H2RG detector and hardware and software support from Teledyne Scientific and Imaging. RATIR, the automation of the Harold L. Johnson Telescope of the Observatorio Astron\'{o}mico Nacional on Sierra San Pedro M\'{a}rtir, and the operation of both are funded through NASA grants NNX09AH71G, NNX09AT02G, NNX10AI27G, and NNX12AE66G, CONACyT grants INFR-2009-01-122785 and CB-2008-101958 , UNAM PAPIIT grant IG100414, and UC MEXUS-CONACyT grant CN 09-283.

These results made use of Lowell Observatory's Discovery Channel Telescope.
Lowell operates the DCT in partnership with Boston University, Northern Arizona University, the University of Maryland, and the University of Toledo.  Partial support of the DCT was provided by Discovery Communications.  LMI was built by Lowell Observatory using funds from the NSF (AST-1005313).
The Liverpool Telescope is operated on the island of La Palma by Liverpool John Moores University in the Spanish Observatorio del Roque de los Muchachos of the Instituto de Astrofisica de Canarias with financial support from the UK Science and Technology Facilities Council.
Some of the data presented herein were obtained at the W. M. Keck
Observatory, which is operated as a scientific partnership among the
California Institute of Technology, the University of California, and
NASA; the observatory was made possible by the generous financial
support of the W. M. Keck Foundation.

This research has made use of the VizieR catalogue access tool, CDS, Strasbourg, France.
This publication also uses data products from the Two Micron All Sky Survey, which is a joint project of the University of Massachusetts and the Infrared Processing and Analysis Center/California Institute of Technology, funded by NASA and the NSF.  Funding for SDSS-III has been provided by the Alfred P. Sloan Foundation, the Participating Institutions, the NSF, and the U.S. Department of Energy Office of Science. The SDSS-III website is http://www.sdss3.org/.  SDSS-III is managed by the Astrophysical Research Consortium for the Participating Institutions of the SDSS-III Collaboration including the University of Arizona, the Brazilian Participation Group, Brookhaven National Laboratory, Carnegie Mellon University, University of Florida, the French Participation Group, the German Participation Group, Harvard University, the Instituto de Astrofisica de Canarias, the Michigan State/Notre Dame/JINA Participation Group, Johns Hopkins University, Lawrence Berkeley National Laboratory, Max Planck Institute for Astrophysics, Max Planck Institute for Extraterrestrial Physics, New Mexico State University, New York University, Ohio State University, Pennsylvania State University, University of Portsmouth, Princeton University, the Spanish Participation Group, University of Tokyo, University of Utah, Vanderbilt University, University of Virginia, University of Washington, and Yale University. 

\bibliographystyle{apj}
\bibliography{grb130702a}

\clearpage

\LongTables
\begin{center}
\begin{deluxetable*}{ccccc}
\tabletypesize{\footnotesize}
\tablecolumns{12}
\tablewidth{0pt}
\tablecaption{Imaging log}
\label{tab:imaging}
\tablehead{
\colhead{Filter} & \colhead{Epoch} & \colhead{Telescope} & \colhead{Exp.} & \colhead{AB mag}   \\
& \colhead{(days)} & & \colhead{(s)} &
}

\startdata
$u^{\prime}$    &   330.49 &   Keck/LRIS &  203 &  24.50 $\pm$ 0.14 \\
$g^{\prime}$	&	1.17	&	P60	&	120	&	18.80 $\pm$ 0.04	\\
$g^{\prime}$	&	1.26	&	P60	&	120	&	18.86 $\pm$ 0.04	\\
$g^{\prime}$	&	2.21	&	P60	&	270	&	19.52 $\pm$ 0.04	\\
$g^{\prime}$	&	3.23	&	P60	&	450	&	20.02 $\pm$ 0.05	\\
$g^{\prime}$	&	4.19	&	P60	&	1260	&	20.22 $\pm$ 0.04	\\
$g^{\prime}$	&	5.27	&	P60	&	540	&	20.35 $\pm$ 0.04	\\
$g^{\prime}$	&	6.28	&	P60	&	540	&	20.31 $\pm$ 0.04	\\
$g^{\prime}$	&	7.20	&	P60	&	720	&	20.32 $\pm$ 0.04	\\
$g^{\prime}$	&	9.87	&	Liverpool	&	130	&	20.35 $\pm$ 0.33	\\
$g^{\prime}$	&	10.87	&	Liverpool	&	130	&	20.39 $\pm$ 0.35	\\
$g^{\prime}$	&	11.26	&	P60	&	720	&	20.22 $\pm$ 0.05	\\
$g^{\prime}$	&	11.87	&	Liverpool	&	130	&	20.06 $\pm$ 0.21	\\
$g^{\prime}$	&	12.18	&	P60	&	660	&	20.23 $\pm$ 0.03	\\
$g^{\prime}$	&	12.86	&	Liverpool	&	130	&	20.00 $\pm$ 0.23	\\
$g^{\prime}$	&	13.86	&	Liverpool	&	130	&	20.39 $\pm$ 0.34	\\
$g^{\prime}$	&	14.18	&	P60	&	720	&	20.31 $\pm$ 0.06	\\
$g^{\prime}$	&	15.18	&	P60	&	720	&	20.43 $\pm$ 0.07	\\
$g^{\prime}$	&	16.18	&	P60	&	720	&	20.36 $\pm$ 0.13	\\
$g^{\prime}$	&	17.87	&	Liverpool	&	150	&	20.32 $\pm$ 0.13	\\
$g^{\prime}$	&	19.87	&	Liverpool	&	150	&	20.78 $\pm$ 0.19	\\
$g^{\prime}$	&	21.87	&	Liverpool	&	150	&	20.98 $\pm$ 0.15	\\
$g^{\prime}$	&	23.87	&	Liverpool	&	150	&	21.25 $\pm$ 0.30	\\
$g^{\prime}$	&	27.92	&	Liverpool	&	150	&	21.53 $\pm$ 0.07	\\
$g^{\prime}$	&	29.21	&	P60	&	540	&	21.62 $\pm$ 0.12	\\
$g^{\prime}$	&	30.19	&	P60	&	1080	&	21.72 $\pm$ 0.09	\\
$g^{\prime}$	&	34.17	&	P60	&	1080	&	21.84 $\pm$ 0.08	\\
$g^{\prime}$	&	37.89	&	Liverpool	&	450	&	22.06 $\pm$ 0.08	\\
$g^{\prime}$	&	39.88	&	Liverpool	&	450	&	22.11 $\pm$ 0.09	\\
$g^{\prime}$	&	45.88	&	Liverpool	&	450	&	22.64 $\pm$ 0.34	\\
$g^{\prime}$	&	53.88	&	Liverpool	&	150	&	22.42 $\pm$ 0.07	\\
$g^{\prime}$	&	58.88	&	Liverpool	&	150	&	22.56 $\pm$ 0.06	\\
$g^{\prime}$	&	62.87	&	Liverpool	&	150	&	22.60 $\pm$ 0.09	\\
$g^{\prime}$    &   330.48 &   Keck/LRIS   &   203 &   23.51 $\pm$ 0.03    \\
$g^{\prime}$	&	632.37	&	DCT	&	1200	&	23.66 $\pm$ 0.04	\\
$r^{\prime}$	&	0.18	&	P48	&	60	&	17.38 $\pm$ 0.04	\\
$r^{\prime}$	&	0.21	&	P48	&	60	&	17.52 $\pm$ 0.04	\\
$r^{\prime}$	&	1.26	&	P60	&	360	&	18.66 $\pm$ 0.05	\\
$r^{\prime}$	&	2.16	&	RATIR	&	7600	&	19.23 $\pm$ 0.06	\\
$r^{\prime}$	&	2.21	&	P60	&	420	&	19.32 $\pm$ 0.04	\\
$r^{\prime}$	&	3.21	&	RATIR	&	5040	&	19.67 $\pm$ 0.04	\\
$r^{\prime}$	&	3.25	&	P60	&	120	&	19.70 $\pm$ 0.08	\\
$r^{\prime}$	&	4.18	&	P60	&	1320	&	19.97 $\pm$ 0.04	\\
$r^{\prime}$	&	4.21	&	RATIR	&	5360	&	19.88 $\pm$ 0.05	\\
$r^{\prime}$	&	5.21	&	RATIR	&	4960	&	19.95 $\pm$ 0.05	\\
$r^{\prime}$	&	5.27	&	P60	&	540	&	20.00 $\pm$ 0.04	\\
$r^{\prime}$	&	6.20	&	RATIR	&	4880	&	19.94 $\pm$ 0.05	\\
$r^{\prime}$	&	6.27	&	P60	&	540	&	20.04 $\pm$ 0.04	\\
$r^{\prime}$	&	7.19	&	RATIR	&	2640	&	19.90 $\pm$ 0.05	\\
$r^{\prime}$	&	7.19	&	P60	&	720	&	20.06 $\pm$ 0.04	\\
$r^{\prime}$	&	8.22	&	RATIR	&	3920	&	19.82 $\pm$ 0.05	\\
$r^{\prime}$	&	9.87	&	Liverpool	&	130	&	19.86 $\pm$ 0.07	\\
$r^{\prime}$	&	10.87	&	Liverpool	&	130	&	19.69 $\pm$ 0.07	\\
$r^{\prime}$	&	11.25	&	P60	&	660	&	19.82 $\pm$ 0.02	\\
$r^{\prime}$	&	11.87	&	Liverpool	&	130	&	19.73 $\pm$ 0.07	\\
$r^{\prime}$	&	12.18	&	P60	&	660	&	19.78 $\pm$ 0.04	\\
$r^{\prime}$	&	12.22	&	RATIR	&	2880	&	19.70 $\pm$ 0.04	\\
$r^{\prime}$	&	12.87	&	Liverpool	&	130	&	19.74 $\pm$ 0.08	\\
$r^{\prime}$	&	13.16	&	RATIR	&	1280	&	19.68 $\pm$ 0.05	\\
$r^{\prime}$	&	13.87	&	Liverpool	&	130	&	19.68 $\pm$ 0.10	\\
$r^{\prime}$	&	14.18	&	P60	&	660	&	19.70 $\pm$ 0.04	\\
$r^{\prime}$	&	14.18	&	RATIR	&	3760	&	19.68 $\pm$ 0.07	\\
$r^{\prime}$	&	15.16	&	RATIR	&	5280	&	19.70 $\pm$ 0.05	\\
$r^{\prime}$	&	15.17	&	P60	&	660	&	19.64 $\pm$ 0.04	\\
$r^{\prime}$	&	15.87	&	Liverpool	&	150	&	19.80 $\pm$ 0.07	\\
$r^{\prime}$	&	16.17	&	P60	&	660	&	19.76 $\pm$ 0.08	\\
$r^{\prime}$	&	17.87	&	Liverpool	&	150	&	19.83 $\pm$ 0.03	\\
$r^{\prime}$	&	19.87	&	Liverpool	&	150	&	19.92 $\pm$ 0.04	\\
$r^{\prime}$	&	21.88	&	Liverpool	&	150	&	20.00 $\pm$ 0.03	\\
$r^{\prime}$	&	23.87	&	Liverpool	&	150	&	20.20 $\pm$ 0.02	\\
$r^{\prime}$	&	25.17	&	RATIR	&	4560	&	20.16 $\pm$ 0.07	\\
$r^{\prime}$	&	26.15	&	RATIR	&	2160	&	20.22 $\pm$ 0.05	\\
$r^{\prime}$	&	27.15	&	RATIR	&	5200	&	20.31 $\pm$ 0.05	\\
$r^{\prime}$	&	27.91	&	Liverpool	&	150	&	20.42 $\pm$ 0.03	\\
$r^{\prime}$	&	28.16	&	RATIR	&	5120	&	20.38 $\pm$ 0.05	\\
$r^{\prime}$	&	29.15	&	RATIR	&	5200	&	20.52 $\pm$ 0.06	\\
$r^{\prime}$	&	29.19	&	P60	&	540	&	20.49 $\pm$ 0.05	\\
$r^{\prime}$	&	30.18	&	P60	&	540	&	20.54 $\pm$ 0.06	\\
$r^{\prime}$	&	30.19	&	RATIR	&	3120	&	20.63 $\pm$ 0.06	\\
$r^{\prime}$	&	31.15	&	RATIR	&	3920	&	20.62 $\pm$ 0.15	\\
$r^{\prime}$	&	32.15	&	RATIR	&	4960	&	20.73 $\pm$ 0.06	\\
$r^{\prime}$	&	34.16	&	P60	&	540	&	20.87 $\pm$ 0.07	\\
$r^{\prime}$	&	35.17	&	RATIR	&	3680	&	20.96 $\pm$ 0.06	\\
$r^{\prime}$	&	37.88	&	Liverpool	&	150	&	21.22 $\pm$ 0.04	\\
$r^{\prime}$	&	39.88	&	Liverpool	&	150	&	21.27 $\pm$ 0.06	\\
$r^{\prime}$	&	41.16	&	RATIR	&	3840	&	21.25 $\pm$ 0.08	\\
$r^{\prime}$	&	45.15	&	RATIR	&	2080	&	21.33 $\pm$ 0.12	\\
$r^{\prime}$	&	45.87	&	Liverpool	&	150	&	21.46 $\pm$ 0.14	\\
$r^{\prime}$	&	46.15	&	RATIR	&	3120	&	21.36 $\pm$ 0.11	\\
$r^{\prime}$	&	47.88	&	Liverpool	&	150	&	21.42 $\pm$ 0.15	\\
$r^{\prime}$	&	53.87	&	Liverpool	&	150	&	21.75 $\pm$ 0.06	\\
$r^{\prime}$	&	58.87	&	Liverpool	&	150	&	21.81 $\pm$ 0.06	\\
$r^{\prime}$	&	62.86	&	Liverpool	&	150	&	21.80 $\pm$ 0.10	\\
$R$             &   330.48 &   Keck/LRIS   &   200 &   23.26 $\pm$ 0.06    \\
$r^{\prime}$	&	632.39	&	DCT	&	1200	&	23.15 $\pm$ 0.04	\\
$i^{\prime}$	&	1.17	&	P60	&	120	&	18.42 $\pm$ 0.04	\\
$i^{\prime}$	&	1.26	&	P60	&	120	&	18.56 $\pm$ 0.06	\\
$i^{\prime}$	&	2.16	&	RATIR	&	7600	&	19.09 $\pm$ 0.07	\\
$i^{\prime}$	&	2.21	&	P60	&	420	&	19.12 $\pm$ 0.04	\\
$i^{\prime}$	&	3.21	&	RATIR	&	5040	&	19.59 $\pm$ 0.06	\\
$i^{\prime}$	&	3.24	&	P60	&	450	&	19.64 $\pm$ 0.05	\\
$i^{\prime}$	&	4.17	&	P60	&	1320	&	19.97 $\pm$ 0.06	\\
$i^{\prime}$	&	4.21	&	RATIR	&	5360	&	19.93 $\pm$ 0.06	\\
$i^{\prime}$	&	5.21	&	RATIR	&	4960	&	20.05 $\pm$ 0.12	\\
$i^{\prime}$	&	5.26	&	P60	&	540	&	20.13 $\pm$ 0.05	\\
$i^{\prime}$	&	6.20	&	RATIR	&	4880	&	20.13 $\pm$ 0.09	\\
$i^{\prime}$	&	6.27	&	P60	&	660	&	20.09 $\pm$ 0.05	\\
$i^{\prime}$	&	7.18	&	P60	&	660	&	20.17 $\pm$ 0.05	\\
$i^{\prime}$	&	7.19	&	RATIR	&	3280	&	20.07 $\pm$ 0.17	\\
$i^{\prime}$	&	8.22	&	RATIR	&	3920	&	20.10 $\pm$ 0.07	\\
$i^{\prime}$	&	9.87	&	Liverpool	&	130	&	19.93 $\pm$ 0.13	\\
$i^{\prime}$	&	10.87	&	Liverpool	&	130	&	19.86 $\pm$ 0.16	\\
$i^{\prime}$	&	11.17	&	P60	&	180	&	19.88 $\pm$ 0.11	\\
$i^{\prime}$	&	11.87	&	Liverpool	&	130	&	19.87 $\pm$ 0.15	\\
$i^{\prime}$	&	12.19	&	P60	&	660	&	19.94 $\pm$ 0.04	\\
$i^{\prime}$	&	12.22	&	RATIR	&	2880	&	19.91 $\pm$ 0.05	\\
$i^{\prime}$	&	12.87	&	Liverpool	&	130	&	19.94 $\pm$ 0.14	\\
$i^{\prime}$	&	13.16	&	RATIR	&	1280	&	19.86 $\pm$ 0.05	\\
$i^{\prime}$	&	13.87	&	Liverpool	&	130	&	20.01 $\pm$ 0.21	\\
$i^{\prime}$	&	14.18	&	RATIR	&	3760	&	19.84 $\pm$ 0.05	\\
$i^{\prime}$	&	14.19	&	P60	&	660	&	19.86 $\pm$ 0.06	\\
$i^{\prime}$	&	15.16	&	RATIR	&	5280	&	19.83 $\pm$ 0.05	\\
$i^{\prime}$	&	15.19	&	P60	&	660	&	19.77 $\pm$ 0.06	\\
$i^{\prime}$	&	15.87	&	Liverpool	&	150	&	19.91 $\pm$ 0.10	\\
$i^{\prime}$	&	16.19	&	P60	&	660	&	20.02 $\pm$ 0.10	\\
$i^{\prime}$	&	17.87	&	Liverpool	&	150	&	19.94 $\pm$ 0.07	\\
$i^{\prime}$	&	19.87	&	Liverpool	&	150	&	19.88 $\pm$ 0.05	\\
$i^{\prime}$	&	21.87	&	Liverpool	&	150	&	19.97 $\pm$ 0.04	\\
$i^{\prime}$	&	23.87	&	Liverpool	&	150	&	20.09 $\pm$ 0.05	\\
$i^{\prime}$	&	25.17	&	RATIR	&	4560	&	20.04 $\pm$ 0.06	\\
$i^{\prime}$	&	26.15	&	RATIR	&	2160	&	20.09 $\pm$ 0.05	\\
$i^{\prime}$	&	27.15	&	RATIR	&	5200	&	20.15 $\pm$ 0.06	\\
$i^{\prime}$	&	27.91	&	Liverpool	&	150	&	20.18 $\pm$ 0.02	\\
$i^{\prime}$	&	28.16	&	RATIR	&	5120	&	20.21 $\pm$ 0.05	\\
$i^{\prime}$	&	29.15	&	RATIR	&	5200	&	20.30 $\pm$ 0.07	\\
$i^{\prime}$	&	29.22	&	P60	&	540	&	20.31 $\pm$ 0.06	\\
$i^{\prime}$	&	30.19	&	RATIR	&	3120	&	20.41 $\pm$ 0.07	\\
$i^{\prime}$	&	30.20	&	P60	&	540	&	20.44 $\pm$ 0.08	\\
$i^{\prime}$	&	31.15	&	RATIR	&	4000	&	20.43 $\pm$ 0.14	\\
$i^{\prime}$	&	32.15	&	RATIR	&	5440	&	20.52 $\pm$ 0.07	\\
$i^{\prime}$	&	34.18	&	P60	&	540	&	20.67 $\pm$ 0.07	\\
$i^{\prime}$	&	35.17	&	RATIR	&	3680	&	20.68 $\pm$ 0.07	\\
$i^{\prime}$	&	37.88	&	Liverpool	&	150	&	21.01 $\pm$ 0.05	\\
$i^{\prime}$	&	39.87	&	Liverpool	&	150	&	21.03 $\pm$ 0.08	\\
$i^{\prime}$	&	41.16	&	RATIR	&	3840	&	20.88 $\pm$ 0.08	\\
$i^{\prime}$	&	45.15	&	RATIR	&	2080	&	21.05 $\pm$ 0.11	\\
$i^{\prime}$	&	45.87	&	Liverpool	&	150	&	21.16 $\pm$ 0.10	\\
$i^{\prime}$	&	46.15	&	RATIR	&	3120	&	21.12 $\pm$ 0.10	\\
$i^{\prime}$	&	47.88	&	Liverpool	&	150	&	21.14 $\pm$ 0.13	\\
$i^{\prime}$	&	53.86	&	Liverpool	&	150	&	21.56 $\pm$ 0.09	\\
$i^{\prime}$	&	58.86	&	Liverpool	&	150	&	21.65 $\pm$ 0.09	\\
$i^{\prime}$	&	62.86	&	Liverpool	&	150	&	21.53 $\pm$ 0.12	\\
$i^{\prime}$	&	632.39	&	DCT	&	1200	&	23.03 $\pm$ 0.06	\\
$z^{\prime}$	&	2.16	&	RATIR	&	5700	&	18.91 $\pm$ 0.09	\\
$z^{\prime}$	&	2.21	&	P60	&	870	&	19.02 $\pm$ 0.08	\\
$z^{\prime}$	&	3.21	&	RATIR	&	3780	&	19.36 $\pm$ 0.07	\\
$z^{\prime}$	&	3.24	&	P60	&	840	&	19.43 $\pm$ 0.09	\\
$z^{\prime}$	&	4.19	&	P60	&	1320	&	19.53 $\pm$ 0.10	\\
$z^{\prime}$	&	4.21	&	RATIR	&	4020	&	19.72 $\pm$ 0.12	\\
$z^{\prime}$	&	5.20	&	P60	&	1020	&	19.84 $\pm$ 0.12	\\
$z^{\prime}$	&	5.21	&	RATIR	&	3720	&	19.84 $\pm$ 0.08	\\
$z^{\prime}$	&	6.20	&	RATIR	&	3660	&	19.94 $\pm$ 0.11	\\
$z^{\prime}$	&	6.26	&	P60	&	540	&	19.95 $\pm$ 0.12	\\
$z^{\prime}$	&	7.18	&	P60	&	540	&	20.04 $\pm$ 0.11	\\
$z^{\prime}$	&	7.19	&	RATIR	&	2820	&	19.71 $\pm$ 0.35	\\
$z^{\prime}$	&	8.22	&	RATIR	&	2940	&	19.98 $\pm$ 0.13	\\
$z^{\prime}$	&	9.87	&	Liverpool	&	150	&	20.11 $\pm$ 0.12	\\
$z^{\prime}$	&	10.87	&	Liverpool	&	150	&	19.99 $\pm$ 0.12	\\
$z^{\prime}$	&	11.87	&	Liverpool	&	150	&	20.13 $\pm$ 0.11	\\
$z^{\prime}$	&	12.22	&	RATIR	&	2160	&	19.94 $\pm$ 0.13	\\
$z^{\prime}$	&	12.87	&	Liverpool	&	150	&	20.02 $\pm$ 0.11	\\
$z^{\prime}$	&	13.16	&	RATIR	&	960	&	20.02 $\pm$ 0.12	\\
$z^{\prime}$	&	13.87	&	Liverpool	&	150	&	19.82 $\pm$ 0.13	\\
$z^{\prime}$	&	14.18	&	RATIR	&	2820	&	19.87 $\pm$ 0.12	\\
$z^{\prime}$	&	15.16	&	RATIR	&	3960	&	20.01 $\pm$ 0.10	\\
$z^{\prime}$	&	15.87	&	Liverpool	&	150	&	20.21 $\pm$ 0.20	\\
$z^{\prime}$	&	17.88	&	Liverpool	&	150	&	20.10 $\pm$ 0.08	\\
$z^{\prime}$	&	19.88	&	Liverpool	&	150	&	20.08 $\pm$ 0.05	\\
$z^{\prime}$	&	21.88	&	Liverpool	&	150	&	20.19 $\pm$ 0.06	\\
$z^{\prime}$	&	23.88	&	Liverpool	&	150	&	20.27 $\pm$ 0.04	\\
$z^{\prime}$	&	25.17	&	RATIR	&	3300	&	20.46 $\pm$ 0.14	\\
$z^{\prime}$	&	26.15	&	RATIR	&	1560	&	20.06 $\pm$ 0.13	\\
$z^{\prime}$	&	27.15	&	RATIR	&	3900	&	20.30 $\pm$ 0.09	\\
$z^{\prime}$	&	27.90	&	Liverpool	&	150	&	20.33 $\pm$ 0.07	\\
$z^{\prime}$	&	28.16	&	RATIR	&	3780	&	20.23 $\pm$ 0.09	\\
$z^{\prime}$	&	29.15	&	RATIR	&	3900	&	20.49 $\pm$ 0.10	\\
$z^{\prime}$	&	30.19	&	RATIR	&	2340	&	20.44 $\pm$ 0.15	\\
$z^{\prime}$	&	31.15	&	RATIR	&	2940	&	20.45 $\pm$ 0.11	\\
$z^{\prime}$	&	32.15	&	RATIR	&	4080	&	20.81 $\pm$ 0.15	\\
$z^{\prime}$	&	34.15	&	RATIR	&	3660	&	20.87 $\pm$ 0.12	\\
$z^{\prime}$	&	35.17	&	RATIR	&	2820	&	20.96 $\pm$ 0.14	\\
$z^{\prime}$	&	37.87	&	Liverpool	&	150	&	21.15 $\pm$ 0.15	\\
$z^{\prime}$	&	39.87	&	Liverpool	&	150	&	21.33 $\pm$ 0.20	\\
$z^{\prime}$	&	45.86	&	Liverpool	&	150	&	21.41 $\pm$ 0.30	\\
$z^{\prime}$	&	53.85	&	Liverpool	&	150	&	21.27 $\pm$ 0.13	\\
$z^{\prime}$	&	58.85	&	Liverpool	&	150	&	21.60 $\pm$ 0.20	\\
$z^{\prime}$	&	62.85	&	Liverpool	&	150	&	21.85 $\pm$ 0.29	\\
$z^{\prime}$	&	632.41	&	DCT	&	1200	&	23.02 $\pm$ 0.12	\\
$y$	&	2.16	&	RATIR	&	5700	&	18.78 $\pm$ 0.08	\\
$y$	&	3.21	&	RATIR	&	3780	&	19.23 $\pm$ 0.08	\\
$y$	&	4.21	&	RATIR	&	4020	&	19.53 $\pm$ 0.09	\\
$y$	&	5.21	&	RATIR	&	3720	&	19.70 $\pm$ 0.09	\\
$y$	&	6.20	&	RATIR	&	3660	&	19.79 $\pm$ 0.07	\\
$y$	&	12.22	&	RATIR	&	2160	&	19.49 $\pm$ 0.10	\\
$y$	&	13.16	&	RATIR	&	960	&	19.70 $\pm$ 0.13	\\
$y$	&	14.18	&	RATIR	&	2820	&	19.75 $\pm$ 0.09	\\
$y$	&	15.16	&	RATIR	&	3960	&	19.73 $\pm$ 0.11	\\
$y$	&	22.16	&	RATIR	&	2340	&	19.85 $\pm$ 0.13	\\
$y$	&	25.17	&	RATIR	&	3300	&	19.69 $\pm$ 0.14	\\
$y$	&	26.15	&	RATIR	&	1560	&	20.00 $\pm$ 0.13	\\
$y$	&	27.15	&	RATIR	&	3900	&	20.05 $\pm$ 0.09	\\
$y$	&	28.16	&	RATIR	&	3780	&	20.01 $\pm$ 0.09	\\
$y$	&	29.15	&	RATIR	&	3900	&	20.03 $\pm$ 0.10	\\
$y$	&	30.19	&	RATIR	&	2340	&	20.04 $\pm$ 0.11	\\
$y$	&	31.15	&	RATIR	&	2940	&	20.28 $\pm$ 0.12	\\
$y$	&	32.15	&	RATIR	&	4080	&	20.00 $\pm$ 0.11	\\
$y$	&	34.15	&	RATIR	&	3660	&	20.49 $\pm$ 0.14	\\
$y$	&	35.17	&	RATIR	&	2820	&	20.51 $\pm$ 0.14	\\
$J$	&	2.16	&	RATIR	&	6380	&	18.75 $\pm$ 0.09	\\
$J$	&	3.21	&	RATIR	&	4230	&	19.22 $\pm$ 0.07	\\
$J$	&	4.21	&	RATIR	&	4500	&	19.65 $\pm$ 0.09	\\
$J$	&	5.21	&	RATIR	&	4160	&	19.64 $\pm$ 0.10	\\
$J$	&	6.20	&	RATIR	&	4090	&	19.85 $\pm$ 0.10	\\
$J$	&	14.18	&	RATIR	&	3150	&	20.07 $\pm$ 0.12	\\
$J$	&	15.16	&	RATIR	&	4430	&	19.86 $\pm$ 0.11	\\
$J$	&	27.15	&	RATIR	&	4360	&	19.92 $\pm$ 0.13	\\
$J$	&	28.16	&	RATIR	&	4230	&	20.05 $\pm$ 0.12	\\
$J$	&	29.15	&	RATIR	&	4360	&	20.23 $\pm$ 0.13	\\
$J$ &  349.45   &  Keck/MOSFIRE & 132   &   23.18 $\pm$ 0.32    \\
$H$	&	2.16	&	RATIR	&	6380	&	18.60 $\pm$ 0.09	\\
$H$	&	3.21	&	RATIR	&	4230	&	18.96 $\pm$ 0.08	\\
$H$	&	4.21	&	RATIR	&	4500	&	19.37 $\pm$ 0.11	\\
$H$	&	5.21	&	RATIR	&	4160	&	19.83 $\pm$ 0.12	\\
$H$	&	6.20	&	RATIR	&	4090	&	19.89 $\pm$ 0.13	\\
$K_{s}$ &  349.44  &  Keck/MOSFIRE &  79  &  $> 22.29$     
\enddata
\tablecomments{AB magnitudes, not corrected for Galactic extinction.}
\label{tab:phot}
\end{deluxetable*}
\end{center}

\end{document}